\documentclass[aps,pra,twocolumn,superscriptaddress,showpacs,aps]{revtex4}
\usepackage{amsmath,amsfonts,amssymb}

\usepackage{graphicx}

\def\sech{{\rm sech}}
\def\eps{\varepsilon}

\begin{document}

\title{Vector nematicons}

\author{Theodoros P. Horikis}
\affiliation{Department of Mathematics, University of Ioannina, Ioannina 45110,
Greece}

\author{Dimitrios J. Frantzeskakis}
\affiliation{Department of Physics, University of Athens, Panepistimiopolis, Zografos,
Athens 15784, Greece}

\begin{abstract}
Families of soliton pairs, namely vector solitons, are found within the context of a coupled
nonlocal nonlinear Schr\"odinger system of equations, as appropriate for modeling beam
propagation in nematic liquid crystals. In the focusing case, bright soliton pairs have been
found to exist provided their amplitudes satisfy a specific condition. In our analytical
approach, focused on the defocusing regime, we rely on a multiscale expansion methods, which
reveals the existence of dark-dark and antidark-antidark solitons, obeying an effective
Korteweg-de Vries equation, as well as dark-bright solitons, obeying an effective Mel'nikov
system. These pairs are discriminated by the sign of a constant that links all physical
parameters of the system to the amplitude of the stable continuous wave solutions, and, much like
the focusing case, the solitons' amplitudes are linked leading to mutual guiding.
\end{abstract}


\maketitle

\section{Introduction}

Solitons, namely robust localized waveforms propagating in nonlinear dispersive/diffractive media,
have been studied extensively in various physical contexts \cite{daux}. In nonlinear optics,
solitons appear either as pulses localized in time (temporal solitons) or as bounded self-guided
beams in space (spatial solitons) \cite{kivshar_book}. These structures are usually described by
the two main variants of the nonlinear Schr\"odinger (NLS) equation, with a local Kerr (cubic)
nonlinearity and depend on the relative sign of dispersion/diffraction and nonlinearity: the
focusing, where dispersion/diffraction and nonlinearity share the same signs and bright solitons
are exhibited and defocusing where the two effects have opposite signs and the NLS supports dark
solitons. In the case where more than one-component scalar fields are involved (as in the case of
fields of different frequencies or different polarizations), their nonlinear interaction leads to
vector NLS models, which support {\it vector solitons} of various types, e.g., bright-bright,
dark-dark, dark-bright, and so on, depending again on the relative signs of dispersion/diffraction,
as well as inter- and intra-component nonlinearity coefficients \cite{kivshar_book,ofy}. Note that
a similar picture, regarding vector NLS models with local cubic nonlinearities and the types of
vector solitons they support, appear in other physical systems, such as atomic Bose-Einstein
condensates \cite{siam,revip}.

On the other hand, there has been an increased interest in physical systems (and their
corresponding mathematical models) featuring a spatially nonlocal nonlinear response, where beam
dynamics and solitons are relevant. Pertinent examples include partially ionized plasmas
\cite{litvak,plasma}, atomic vapors \cite{vap}, lead glasses featuring strong thermal nonlinearity
\cite{rot}, as well as media with a long-range inter-particle interaction. The latter include
dipolar bosonic quantum gases \cite{santos}, and nematic liquid crystals with long-range molecular
reorientational interactions \cite{ass0}. Nematic liquid crystals are known to support spatial
solitons \cite{assanto}, which are usually called {\it nematicons}
\cite{ass2,assanto_book,peccianti}. These structures, are described by nonlocal NLS equations
which, in general, do not possess exact analytical solutions with the freedom of various
parameters describing the soliton's properties (amplitude, velocity, etc). Thus, variational
techniques are usually employed for the study of either bright
\cite{worthy2,skuse3,alberucci,assanto3} or dark \cite{karta,piccardi,kong,assanto_grey,pu}
nematicons. More recently, in the self-defocusing setting, multiscale expansion methods were used
to study dark nematicons in one-dimensional (1D) \cite{small_amplitude} and higher-dimensional
\cite{djfth,djfth2} geometries; these studies, apart from investigating the dynamics of dark
solitons, also predicted the existence of {\it antidark} solitons, namely humps on top of a
continuous-wave (cw) background. These solutions are discriminated from dark solitons by the sign
of a specific parameter, which associates the degree of nonlocality with the amplitude of the cw
wave on top of which these solutions are formed.

Two-color nematicons, i.e. vectorial nematicon structures excited at different wavelengths, have
also been experimentally realized and studied theoretically as well \cite{vector,worthy,skuse3,xu}.
In the focusing 1D setting, the existence of exact bright-bright soliton solutions, provided that
their amplitudes satisfy a specific condition, was recently reported \cite{tph_pla}. In the same 1D
setting, but in the defocusing regime, nonlocal dark-dark \cite{kong1} and dark-bright \cite{lin1}
solitons were studied by means of variational methods, similar to those used in the one-component
problem. Notice that the defocusing regime is also accessible in the context of nematic liquid
crystals: indeed, as shown in Ref.~\cite{piccardi} where dark nematicons were observed for the
first time, azo-doped nematic liquid crystals exhibit a self-defocusing response for extraordinary
waves. Generally, instead of exploiting a thermo-optic response, self-defocusing in this setting
can be obtained by introducing dopants \cite{khoo}.

In this work, our aim is to present families of vector nematicons in the 1D, self-defocusing
setting. Our main findings, as well as the outline of the paper, are as follows. First, in
Section~II, we present the model, as well as review its cw solution, its stability and the
derivation of the appropriate condition for bright solitons to exist. Note that the complete
analysis for this case was presented in Ref. \cite{tph_pla}, we briefly summarize these findings
for completeness. Then, in Section~III, seeking solutions that feature nontrivial boundary
conditions at infinity, we develop a multiscale expansion method that reduces the nonlocal system
to a single Korteweg-de Vries (KdV) equation; we also obtain an additional equation that links the
amplitudes of the two modes, and thus derive dark-dark and antidark-antidark soliton pairs. In
Section~IV, assuming that one mode decays to zero at infinity, we develop another multiscale
expansion method to reduce the nonlocal system to the Mel'nikov system \cite{mel1,mel2}; this
system, which is completely integrable by means of the inverse scattering transform \cite{mel3},
allows for the derivation of dark-bright soliton solutions in the original nonlocal system. In all
cases, our analytical findings are corroborated by direct numerical simulations. Finally, in
Section~V, we summarize our findings and suggest further generalizations.

\section{The governing equations}

We consider the equations that describe two polarised, coherent light beams, of two
different wavelengths, propagating through a cell filled with a nematic liquid
crystal.
These equations are expressed in dimensionless form as follows \cite{vector,skuse2}:
\begin{subequations}
\begin{gather}
i\frac{{\partial {E_1}}}{{\partial z}} + \frac{{{d_1}}}{2}\frac{{{\partial
^2}{E_1}}}{{\partial {x^2}}} + 2 {g_1}\theta
{E_1} = 0, \\
i\frac{{\partial {E_2}}}{{\partial z}} + \frac{{{d_2}}}{2}\frac{{{\partial
^2}{E_2}}}{{\partial {x^2}}} + 2{g_2}\theta
{E_2} = 0,  \\
\nu \frac{{{\partial ^2}\theta }}{{\partial {x^2}}} - 2q\theta  =  -
2({g_1}|{E_1}{|^2}
+ {g_2}|{E_2}{|^2}).
\end{gather}
\label{system11}
\end{subequations}
The variables $E_1$ and $E_2$ are the complex valued, slowly-varying envelopes of the electric
fields, and $\theta$ is the optically induced deviation of the director angle. Diffraction is
characterized by the coefficients $d_1,~d_2$, while nonlinearity by $g_1,~g_2$. The nonlocality
parameter $\nu$ measures the strength of the response of the nematic in space, with a highly
nonlocal response corresponding to $\nu$ large. The parameter $q$ is related to the square of the
applied static field which pre-tilts the nematic dielectric \cite{peccianti,alberucci,assanto3}.
Note that the above system corresponds to the nonlocal regime with $\nu$ large, where the optically
induced rotation $\theta$ is small \cite{assanto3}; in particular, $d_1,g_1,d_2,g_2,q$ are $O(1)$
while $\nu$ is $O(10^2)$ \cite{skuse2,skuse3}. Depending on the relative signs between diffraction
and nonlinearity the relative system is deemed focusing ($d_1g_1,d_2g_2>0$) or defocusing
($d_1g_1,d_2g_2<0$). These equations assume an incoherent interaction between the beams and
that they only interact through the nematic. That is, there are no coupling terms between $E_1$ and
$E_2$.

The simplest solution of this system is a pair of cw's of the form
\[
E_1(z)=u_0 e^{2i g_1 \theta_0 z},\, E_2(z)=v_0 e^{2i g_2 \theta_0 z},\, \theta_0=\frac{g_1
u_0^2+g_2
v_0^2}{q}
\]
where $u_0$ and $v_0$ are real constants. By considering small perturbations to these
solutions in Ref. \cite{tph_pla}, the dispersion relation
\[
p_1(k){\omega ^4} + p_2(k){\omega ^2} + p_3(k)=0
\]
was derived, where
\begin{align*}
p_1(k) &= 16\left( {{k^2}\nu  + 2q} \right)\\
p_2(k) &=  - 4\nu \left( {d_1^2 + d_2^2} \right){k^6} - 8q\left( {d_1^2 + d_2^2}
\right){k^4} \\ &+
64\left( {{d_1}g_1^2u_0^2 + {d_2}g_2^2v_0^2} \right){k^2} \\
p_3(k) &= d_1^2d_2^2\nu {k^{10}} + 2d_1^2d_2^2q{k^8} \\&- 16{d_1}{d_2}\left(
{{d_2}g_1^2u_0^2 +
{d_1}g_2^2v_0^2} \right){k^6}.
\end{align*}
This dispersion relation was shown to have real roots, i.e. the system would be modulationally stable, provided
the diffraction and nonlinearity signs are opposite, i.e. the fully defocusing case.
Hereafter, we fix this sign difference into the nonlocal system and we write
\begin{subequations}
\begin{gather}
i\frac{{\partial {E_1}}}{{\partial z}} + \frac{{{d_1}}}{2}\frac{{{\partial
^2}{E_1}}}{{\partial {x^2}}} - 2{g_1}\theta
{E_1} = 0 \\
i\frac{{\partial {E_2}}}{{\partial z}} + \frac{{{d_2}}}{2}\frac{{{\partial
^2}{E_2}}}{{\partial {x^2}}} - 2{g_2}\theta
{E_2} = 0  \\
\nu \frac{{{\partial ^2}\theta }}{{\partial {x^2}}} - 2q\theta  =  - 2({g_1}|{E_1}{|^2}
+ {g_2}|{E_2}{|^2})
\end{gather}
\label{system1}
\end{subequations}
where now $d_1,g_1,d_2,g_2,\nu,q$ are all positive. Bright soliton pairs of Eqs.
\eqref{system11} have already been discussed in Refs. \cite{tph_pla,mcneil} and will not be
considered here where the focus is turned on the defocusing case.

\section{Dark and antidark soliton pairs}

Our analysis is now focused on soliton pairs that rely on the existence of a stable cw
background and hence on the defocusing system where $d_1g_1,d_2g_2<0$. As such, we only consider Eqs. \eqref{system1}.
Write the solutions of this
system in the form
\begin{subequations}
\begin{eqnarray}
E_1&=&u_b(z) u(z,x),\\
E_2&=&v_b(z) v(z,x),\\
\theta&=&\theta_b w(z,x),
\end{eqnarray}
\label{eq.background}
\end{subequations}
where the functions $u_b(z)$ and $v_b(z)$ correspond to the relative cw backgrounds so
that
\begin{gather*}
\left.\begin{array}{c}
  i{{u'}_b} - 2{g_1}{\theta _b}{u_b} = 0 \\
  i{{v'}_b} - 2{g_2}{\theta _b}{v_b} = 0
\end{array}  \right\} \Rightarrow \left\{ {\begin{array}{c}
  {{u_b}(z) = {u_0}{e^{ - 2i{g_1}{\theta _b}z + i{c_1}}}} \\
  {{v_b}(z) = {v_0}{e^{ - 2i{g_2}{\theta _b}z + i{c_2}}}}
\end{array}} \right.
\label{cws}
\end{gather*}
where $u_0,v_0,c_1,c_2\in\mathbb{R}$ and $\theta_b=\frac{1}{q}(g_1 u_0^2+g_2 v_0^2)$.
Substituting back to Eqs.~\eqref{system1} gives
\begin{subequations}
\begin{gather}
  i\frac{{\partial u}}{{\partial z}} + \frac{{{d_1}}}{2}\frac{{{\partial
  ^2}u}}{{\partial {x^2}}} - 2{g_1}{\theta _b}(w -
  1)u = 0, \\
  i\frac{{\partial v}}{{\partial z}} + \frac{{{d_2}}}{2}\frac{{{\partial
  ^2}v}}{{\partial {x^2}}} - 2{g_2}{\theta _b}(w -
  1)v = 0, \\
  \nu \frac{{{\partial ^2}w}}{{\partial {x^2}}} - 2qw =  - \frac{2}{{{\theta
  _b}}}({g_1}u_0^2|{E_1}{|^2} +
  {g_2}v_0^2|{E_2}{|^2}),
\end{gather}
\label{system2}
\end{subequations}
It is trivial to check that these are also satisfied at the boundaries where
$u=v=w=1$,
and any evolution of the boundary conditions has been absorbed by the background
functions. This way, the resulting equations have now fixed boundary conditions.
Next, we employ the Madelung transformation:
\begin{subequations}
\begin{gather}
u(x,z)=\rho_1(x,z)\exp[i\phi_1(x,z)], \\
v(x,z)=\rho_2(x,z)\exp[i\phi_2(x,z)],
\end{gather}
\end{subequations}
so that:
\begin{subequations}
\begin{gather}
  {d_j}\frac{\partial^2 \rho_{j}}{\partial x^2} - 2{\rho_j}\frac{\partial
  \phi_{j}}{\partial z}
  - {d_j}{\rho_j}\left(\frac{\partial \phi_{j}}{\partial x}\right)^2 -
  4{g_j}{\theta _b}{\rho _j}(w - 1) = 0, \\
  \frac{\partial \rho_j}{\partial z} + \frac{1}{2} d_j \rho_j \frac{\partial^2
  \phi_j}{\partial x^2}
  + {d_j}\frac{\partial \rho_j}{\partial x}\frac{\partial \phi_j}{\partial x}
  = 0, \\
  \nu \frac{\partial^2 w}{\partial x^2} - 2qw =  - \frac{2}{{{\theta
  _b}}}({g_1}u_0^2\rho _1^2 + {g_2}v_0^2\rho
  _2^2),
\end{gather}
\label{madel}
\end{subequations}
where $j=1,2$, and recall that $w(z,x)\in\mathbb{R}$.

To analytically study system~(\ref{madel}), and determine the unknown functions
$\rho_j$, $\phi_j$ and $w$, we now employ the the reductive perturbation method
\cite{Jeffrey}.
We thus introduce the stretched variables:
\begin{equation}
Z=\varepsilon^3 z,\quad X=\varepsilon(x-Cz),
\label{str}
\end{equation}
where $C$ is the speed of sound (to be determined later in the analysis), namely
the velocity of small-amplitude and long-wavelength waves propagating along the
background. Additionally, we expand amplitudes and phases in powers of $\varepsilon$ as
follows:
\begin{subequations}
\begin{eqnarray}
  \rho_j  &=& \rho_{j0} + {\varepsilon ^2}{\rho _{j2}} + {\varepsilon ^4}{\rho _{j4}} +
  \cdots,
  \\
  \phi_j  &=&  \varepsilon {\phi _{j1}}  + {\varepsilon ^3}{\phi _{j3}} + {\varepsilon
  ^5}{\phi _{j5}}+ \cdots, \\
  w&=& 1 + {\varepsilon ^2}{w_2} + {\varepsilon ^4}{w_4} +  \cdots,
\end{eqnarray}
\label{as}
\end{subequations}
\begin{figure*}[ht]
\centering
\includegraphics[height=4cm]{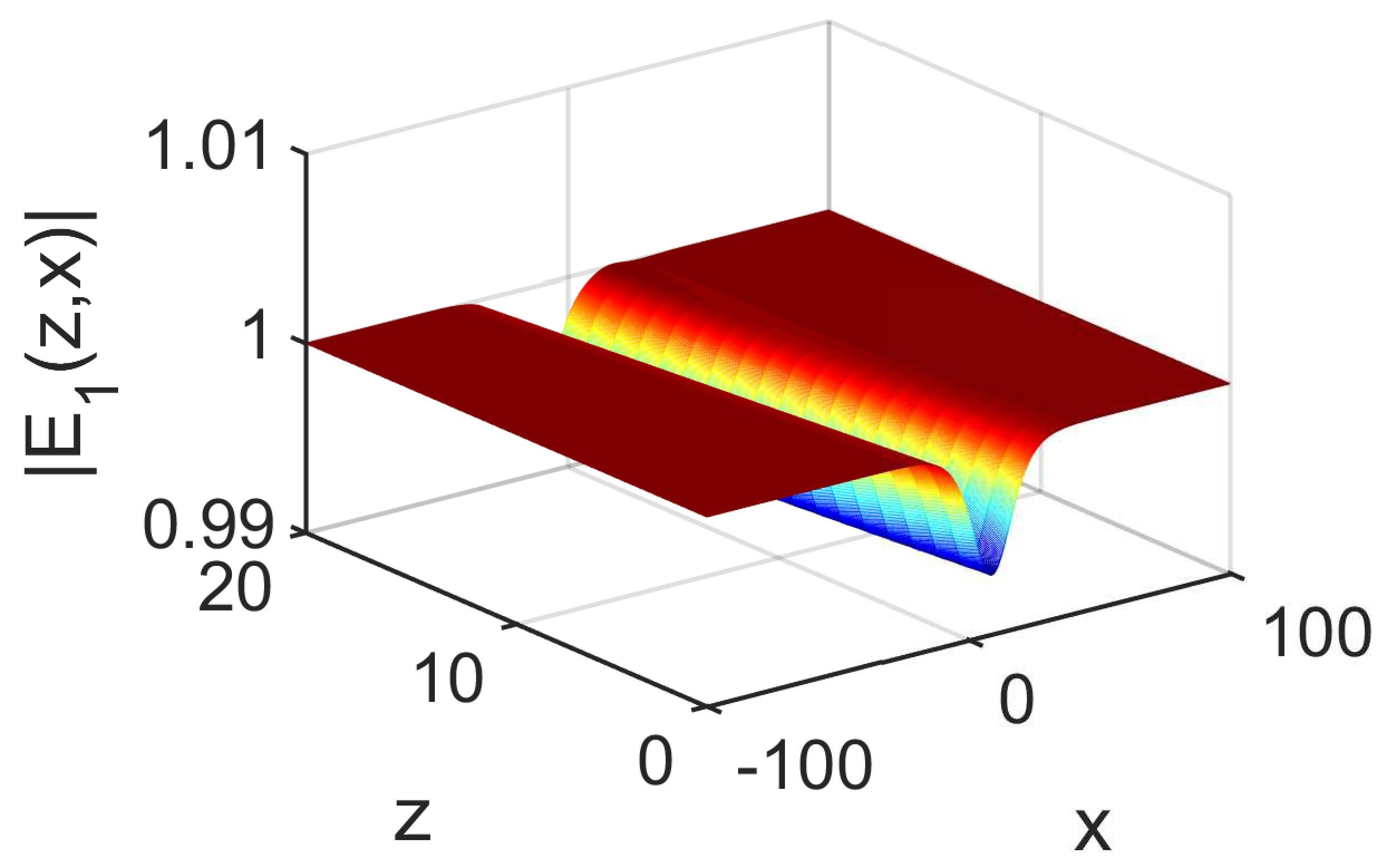}
\includegraphics[height=4cm]{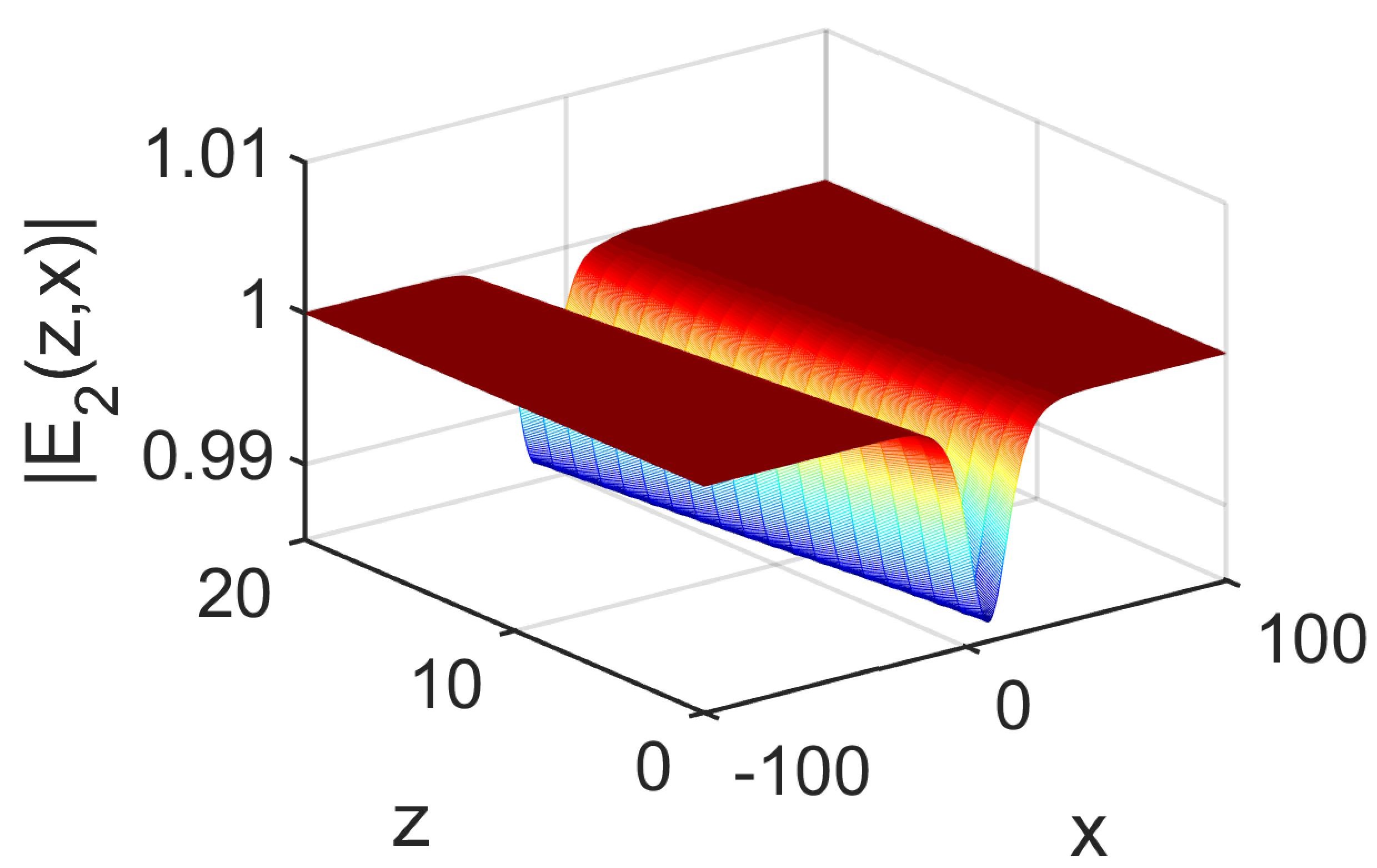}\\
\includegraphics[height=3.5cm]{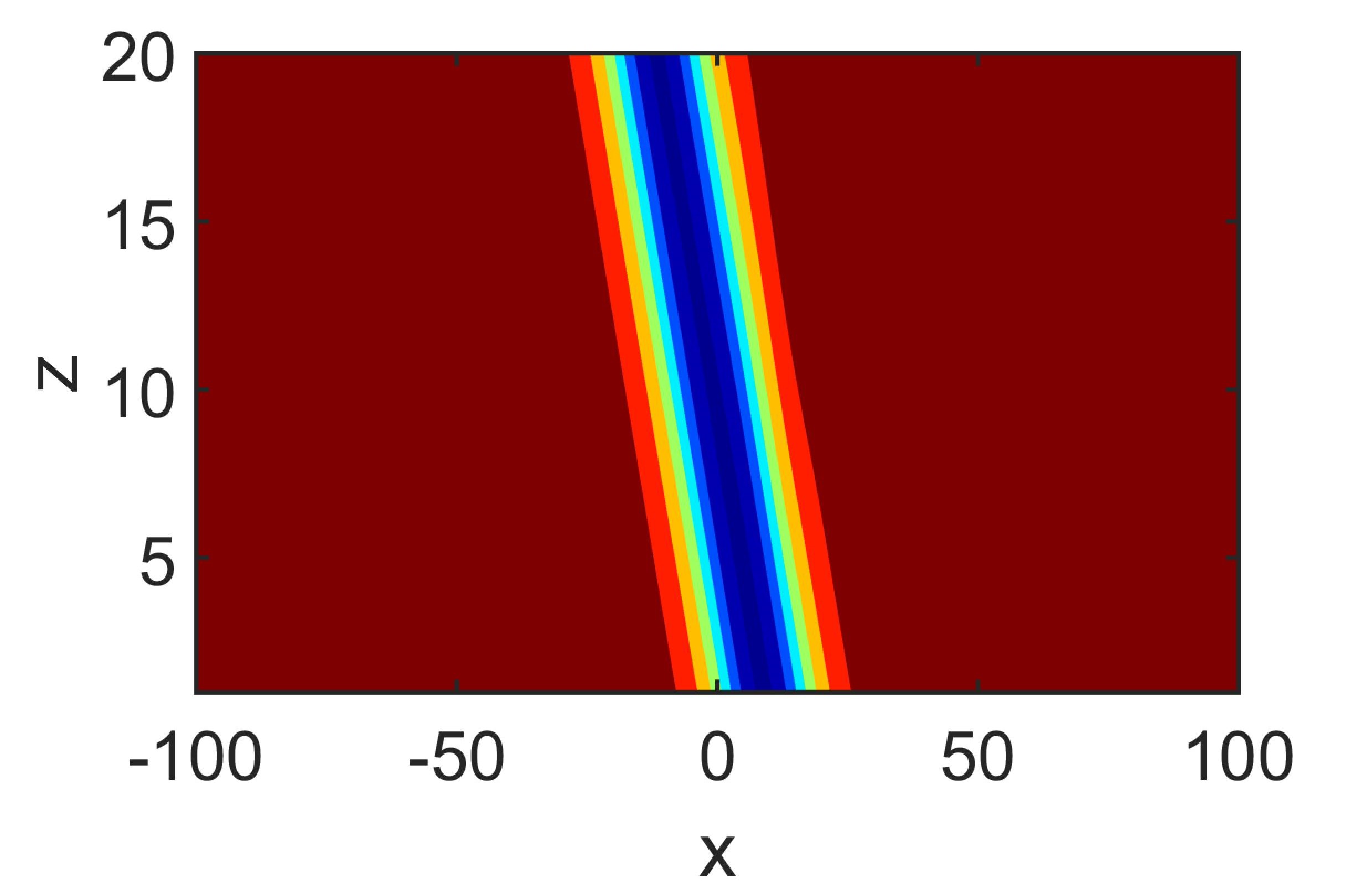}
\includegraphics[height=3.5cm]{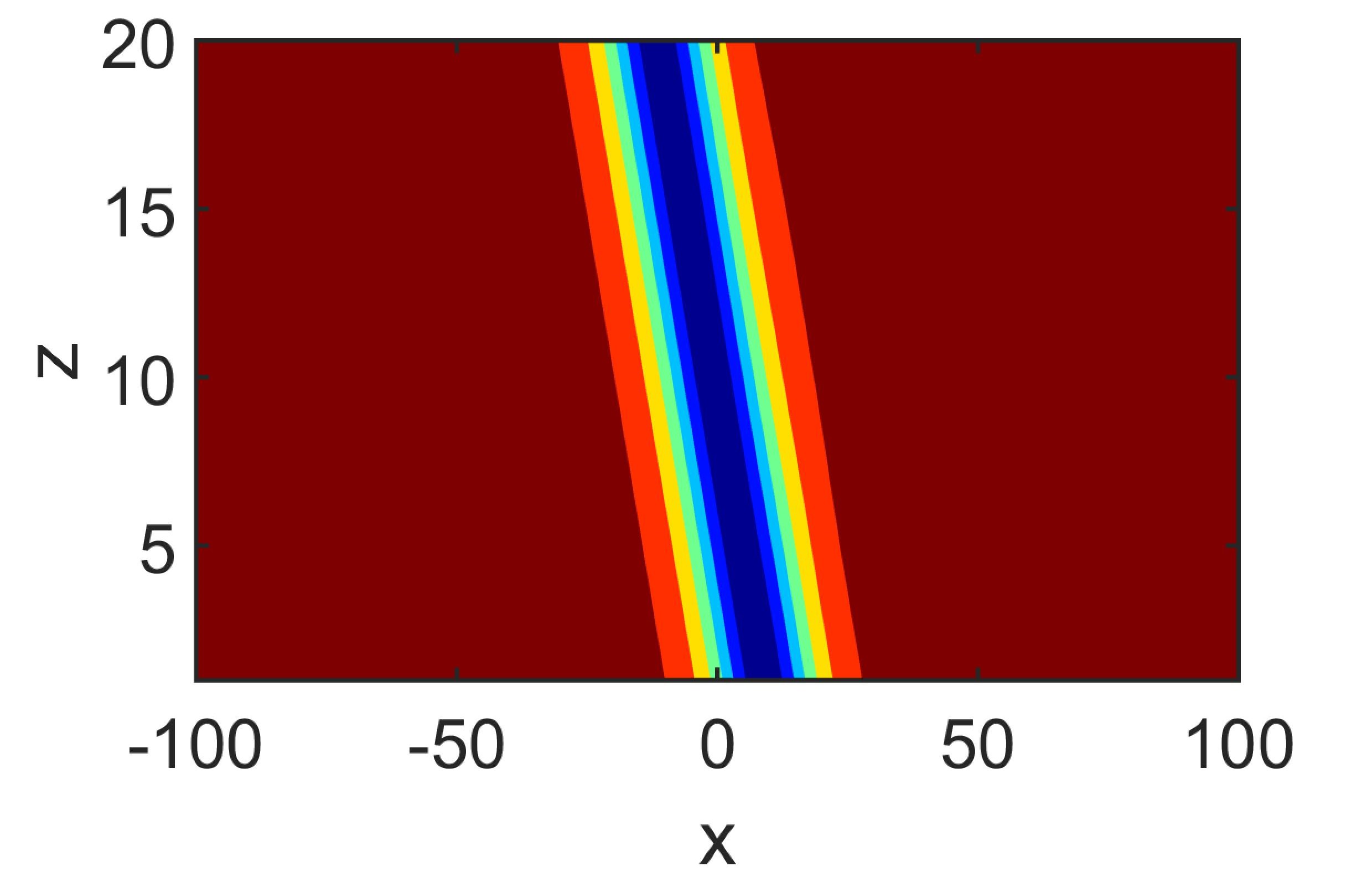}
\caption{(Color Online) The evolution of a typical dark soliton pair. Left and right
columns
depict the two components, while top and bottom panels show three-dimensional plots,
and
spatiotemporal contour plots, respectively. }
\label{dark}
\end{figure*}
\begin{figure*}[ht]
\centering
\includegraphics[height=4cm]{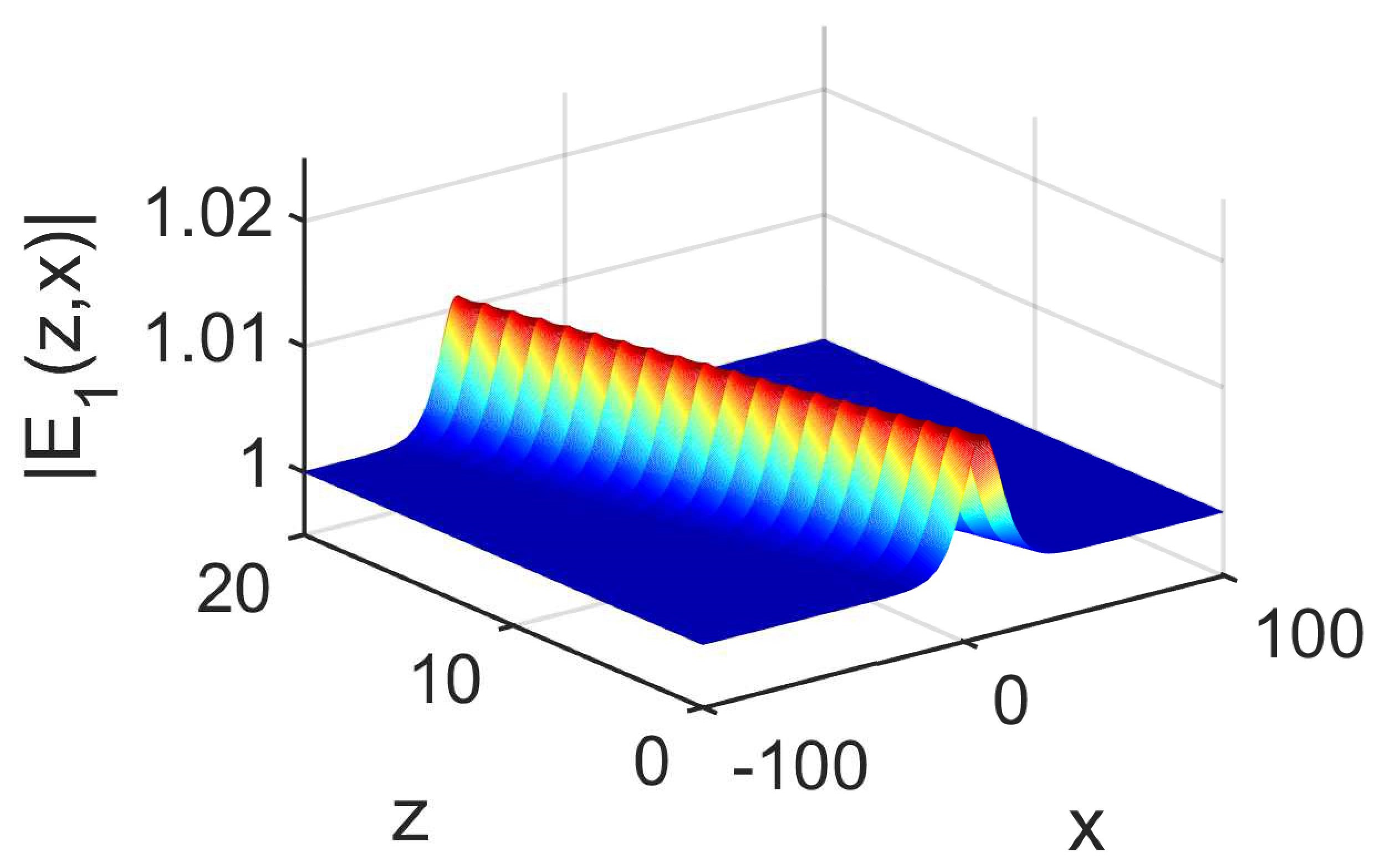}
\includegraphics[height=4cm]{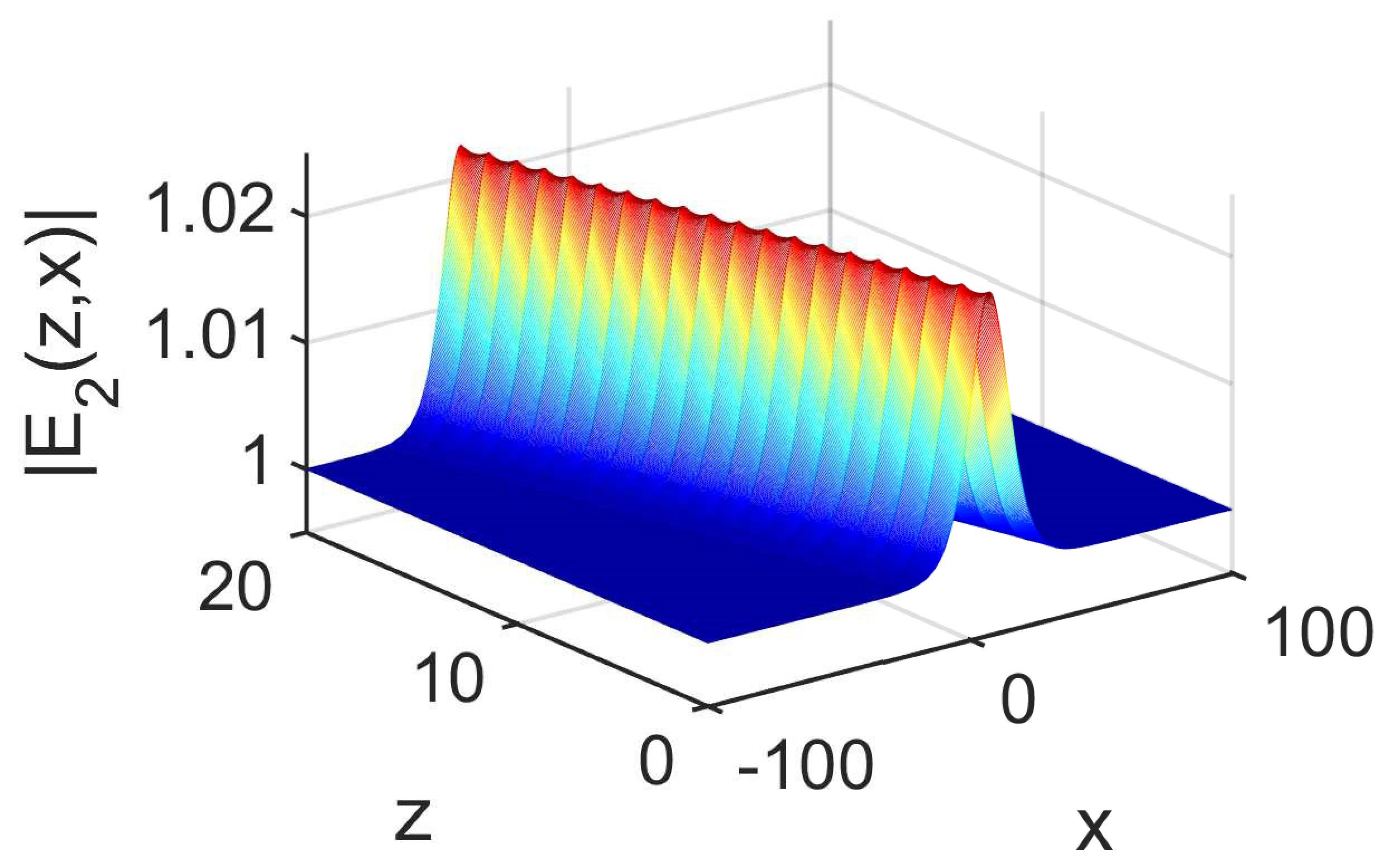}\\
\includegraphics[height=3.5cm]{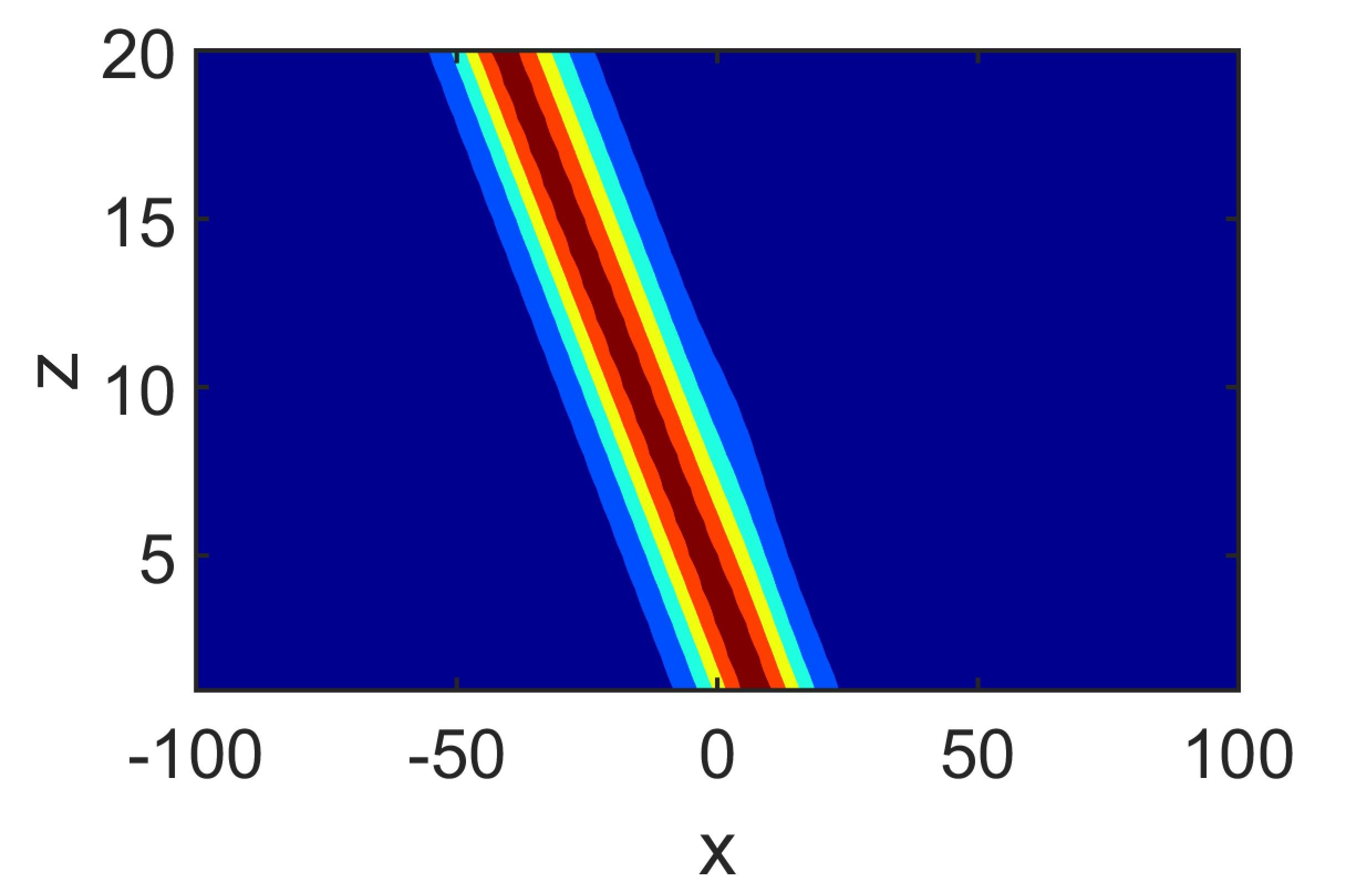}
\includegraphics[height=3.5cm]{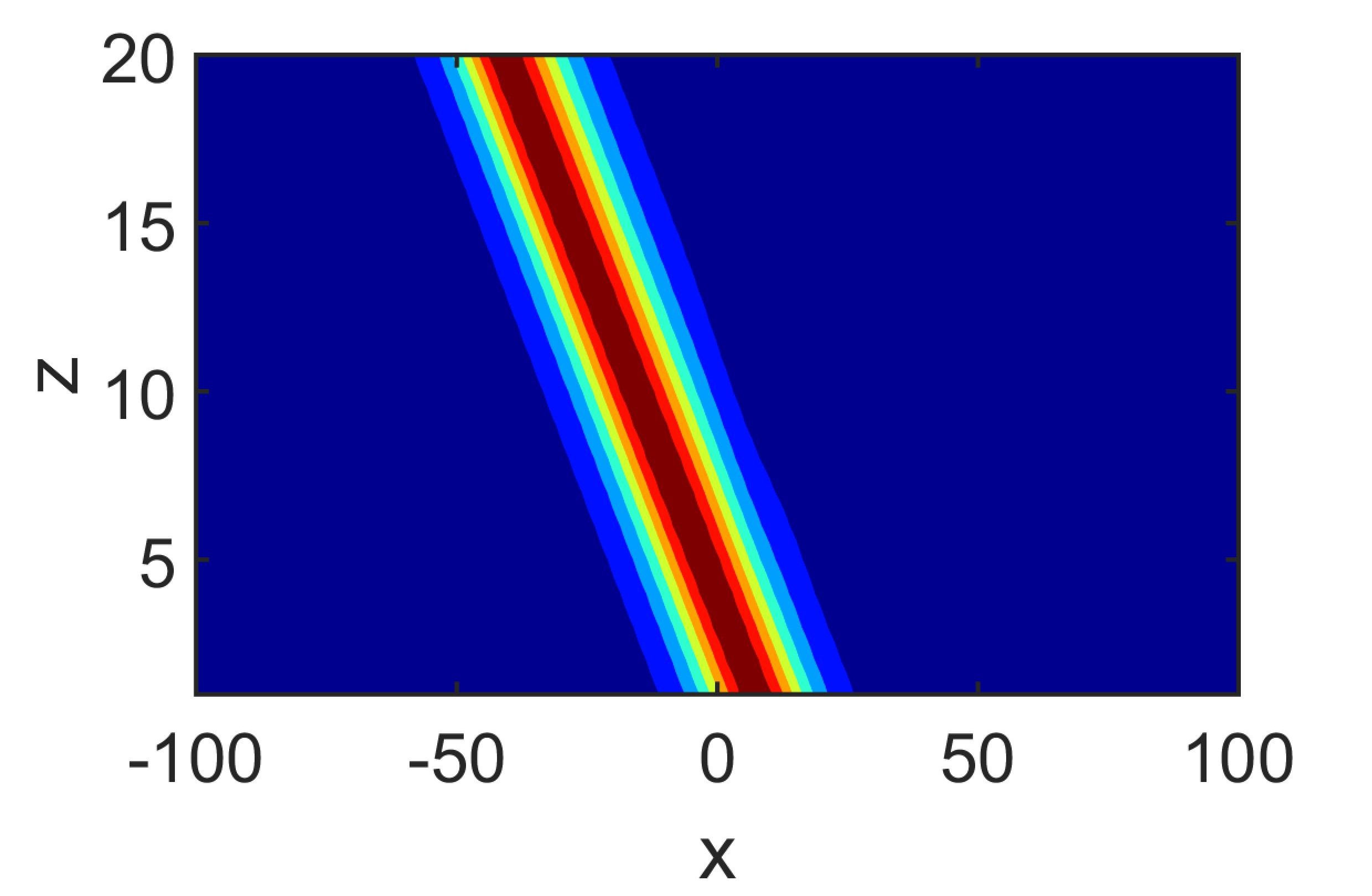}
\caption{(Color Online) Similar to Fig.~\ref{dark}, but now for a typical antidark
soliton pair.}
\label{antidark}
\end{figure*}
where $\rho_{j0}=1$ and the rest of the unknown fields depend on the stretched
variables~(\ref{str}). These values for $\rho_{j0}$ is not only a result obtained from the
perturbation analysis but is also anticipated from Eqs. \eqref{eq.background} and \eqref{system2}.
Recall, that the background has been removed, absorbed by the functions $u_b$ and $v_b$, which, in
general, are not equal.

Substituting back to Eqs.~\eqref{madel} we obtain the following results
(see details in Appendix A). First, in the linear limit, i.e., at the lowest-order
approximation in $\varepsilon$, we derive equations connecting the unknown fields,
namely:
\begin{subequations}
\begin{eqnarray}
w_2 &=& \frac{2}{{q{\theta _b}}}({g_1}u_0^2{\rho _{21}} + {g_2}v_0^2{\rho _{22}}),
\quad
\phi_{21}=\frac{g_2}{g_1}\phi_{11},
\\ \rho_{22}&=&\frac{d_2g_2}{d_1g_1}\rho_{21}, \quad
\frac{d_j}{2}\frac{\partial \phi_{j2}}{\partial X} = C \rho_{j2},
\end{eqnarray}
\label{lin}
\end{subequations}
as well as the speed of sound
\begin{equation}\label{velocity}
C^2 = \frac{2}{q}(d_1g_1u_0^2 + d_2g_2v_0^2).
\end{equation}
Obviously, Eqs.~(\ref{lin}) suggest that only one equation for one of these fields
will suffice to determine the rest of the unknown fields $\rho_{j2}$, $\phi_{j1}$ and
$w_2$.
This equation is derived to the next order of approximation, and turns out to be the
following
nonlinear equation for the field $\rho _{12}$:
\begin{equation}\label{kdv}
\frac{{\partial {\rho _{12}}}}{{\partial Z}} + {A_1}\frac{{{\partial ^3}{\rho
_{12}}}}{{\partial {X^3}}} + 6{A_2}{\rho
_{12}}\frac{{\partial {\rho _{12}}}}{{\partial X}} = 0,
\end{equation}
where coefficients $A_1$ and $A_2$ are given by:
\begin{align*}
{A_1} &= \frac{{\nu {C^4} - (d_1^3g_1^2u_0^2 + d_2^3g_2^2v_0^2)}}{{4{C^2}q}}, \\
A_2 &= \frac{d_1^2g_1^3u_0^2 + d_2^2g_2^3v_0^2}{Cd_1g_1q}.
\end{align*}
Equation~(\ref{kdv}) is the renowned KdV equation, which is completely integrable by means
of the IST
\cite{ist}, and finds numerous applications in a variety of
physical contexts \cite{daux,ablowitz}.
%,Rem,Johnson,Infeld}.
%The KdV equation has also been derived by means of multiscale expansion methods
%from local NLS models, with the aim to describe
%shallow planar dark solitons in Bose gases \cite{tsu} and
%ring dark solitons in nonlinear optical media
%\cite{ofyrds} (see also reviews \cite{ofy,djf} and references therein).
More recently, a KdV equation was derived from the single-component version
of Eqs.~\eqref{system1}, and used to describe small-amplitude
nematicons \cite{small_amplitude}; notice that the KdV model derived in
\cite{small_amplitude} is identical with Eq.~\eqref{kdv} when the coupling constants
are set to zero. Notably, the same procedure can result in other integrable forms of
the KdV in higher dimensions, such as the Kadomtsev-Petviashvilli (KP) equation, Johnson's equation, and others
\cite{djfth,djfth2}.

These asymptotic reductions provide information on the type of the soliton
solutions the original system may exhibit up to (and including) $O(\eps^2)$.
Indeed, first we note that the soliton solution of Eq.~\eqref{kdv} takes
the form (e.g., Ref.~\cite{ablowitz}),
\begin{equation}
\label{soliton}
  \rho_{12}(Z,X)=\frac{2 A_1}{A_2}\eta^2\sech^2(\eta X-4\eta^3 A_1 Z+X_0)
\end{equation}
where $\eta$ and $X_0$ are free parameters, setting the amplitude/width and initial
position of the soliton, respectively. Then, it is straightforward to retrieve
the pertinent phase,
\begin{equation}
\phi_{11}=-\frac{4A_1C}{A_2d_1}\eta\tanh(\eta X-4\eta^3 A_1 Z+X_0),
\end{equation}
so that, finally, the solutions for the two components may be written as:
\begin{eqnarray}
\!\!\!\!\!\!\!\!\!\!\!\!
E_1(z,x) &\approx& u_b(z) (1+\eps^2\rho_{12})\exp(i\eps\phi_{12})
\label{orsol1} \\
\!\!\!\!\!\!\!\!\!\!\!\!
E_2(z,x) &\approx& v_b(z) \left(1+\eps^2\frac{d_2g_2}{d_1g_1}\rho_{12}\right)
\exp\left(i\eps\frac{g_2}{g_1}\phi_{12}\right).
\label{orsol2}
\end{eqnarray}

It is now important to notice that the type of the
solitons~(\ref{orsol1})-(\ref{orsol2})
depends crucially on the sign of the ratio $A_1/A_2$; this quantity changes sign
according to the critical value $\nu_c$, given by:
\begin{gather}
\nu_c=\frac{{{q^2}\left( {d_1^3g_1^2u_0^2 + d_2^3g_2^2v_0^2} \right)}}{{4{{\left(
{{d_1}{g_1}u_0^2 + {d_2}{g_2}v_0^2} \right)}^2}}}.
\end{gather}
Indeed, if the nonlocality parameter $\nu$ is such that $\nu<\nu_c$ (i.e., $A_1/A_2>0$), the
solitons are dark, namely are intensity dips off of the cw background. On the other hand, if
$\nu>\nu_c$ (i.e., $A_1/A_2<0$) the solitons are {\it antidark}, namely intensity elevations on top
of the cw background. Notice that Eqs.~\eqref{amps} suggest that the relative signs between the
modes are the same and, as such, the only allowed pairs are solitons of the same kind. It should
also be mentioned that if $A_1=0$, modification of the asymptotic analysis and inclusion of
higher-order terms is needed. This has been addressed, to a certain extent, in Ref. \cite{el},
where, it was found that higher order dispersive terms can lead to resonant interactions with
radiation, as expected, for the higher (fifth) order KdV equation.

To demonstrate the validity of our analysis, we perform direct numerical simulations
we thus integrate Eqs.~\eqref{system1} employing a high accuracy spectral integrator,
and using initial conditions (at $z=0$) taken from Eqs.~(\ref{orsol1})-(\ref{orsol2}),
for both the dark and the antidark soliton pairs. The results are shown in
Fig.~\ref{dark},
where a typical evolution of a dark soliton pair is depicted.
Here, we choose parameter values $d_1=d_2/1.5=g_1=g_2=1$, $u_0=v_0=1$ and $q/5=\nu=1$.
Similarly, in Fig.~\ref{antidark}, we show a typical evolution of an antidark
soliton pair; all parameters remain the same except $q=1$. In both cases, it is clear
that the solitons, not only exist, but also propagate undistorted on top of the cw
background.
It is also observed that the solitons propagate with constant speed, with the
antidark soliton pair traveling faster than the dark one, as expected from
Eq.~\eqref{velocity}.

\section{Dark-bright soliton pairs}

\begin{figure*}[ht]
\centering
\includegraphics[height=4cm]{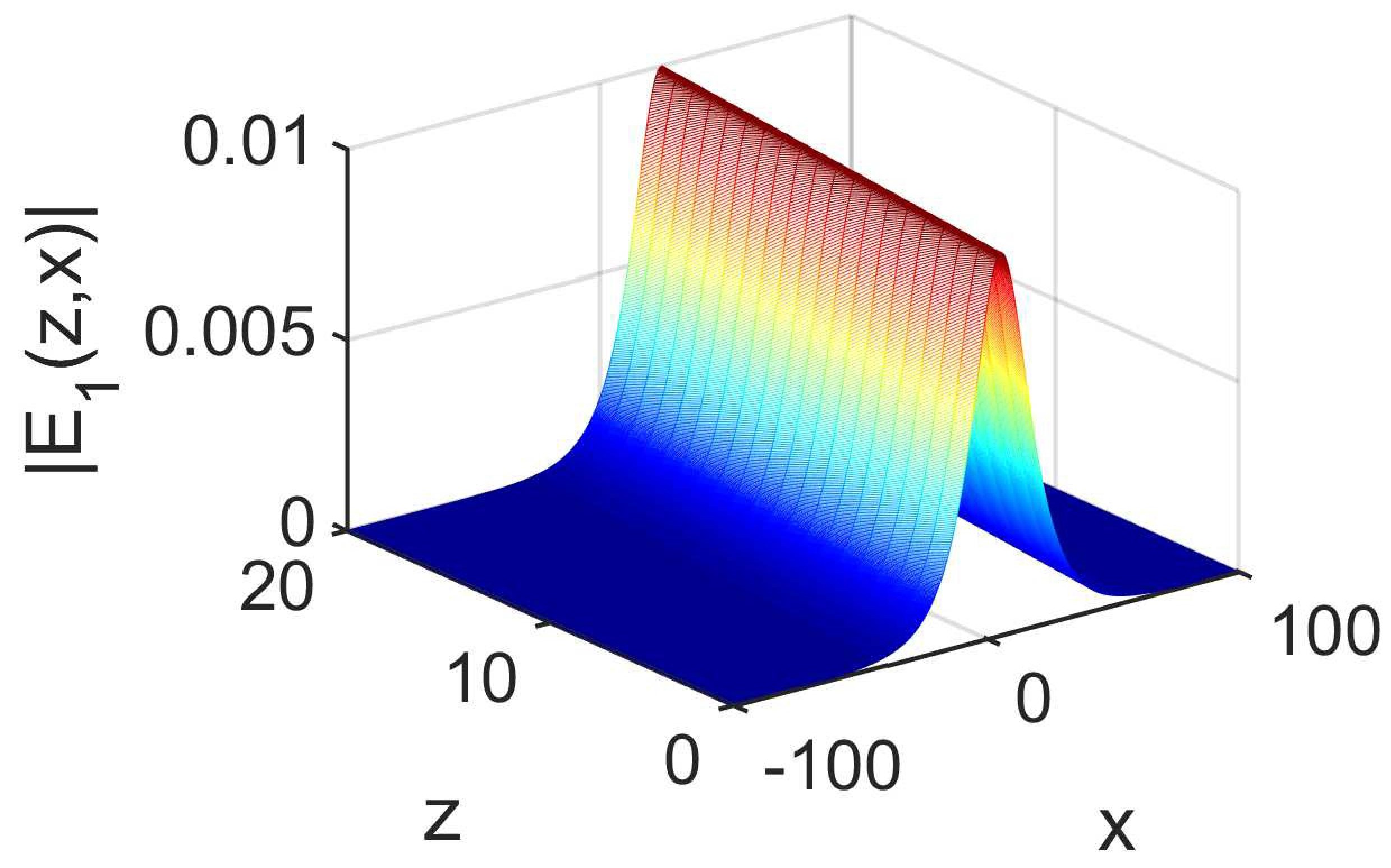}
\includegraphics[height=4cm]{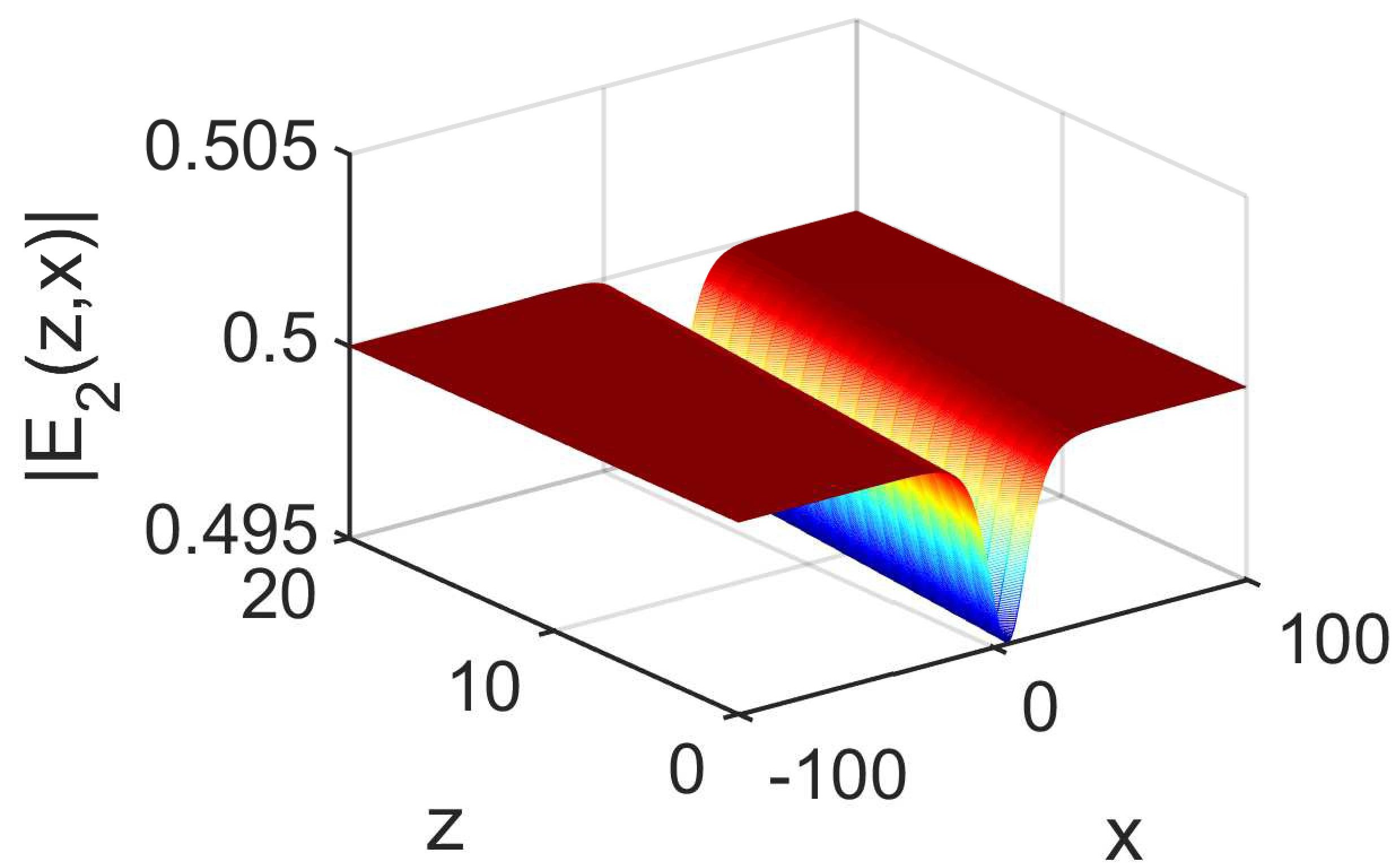} \\
\includegraphics[height=3.5cm]{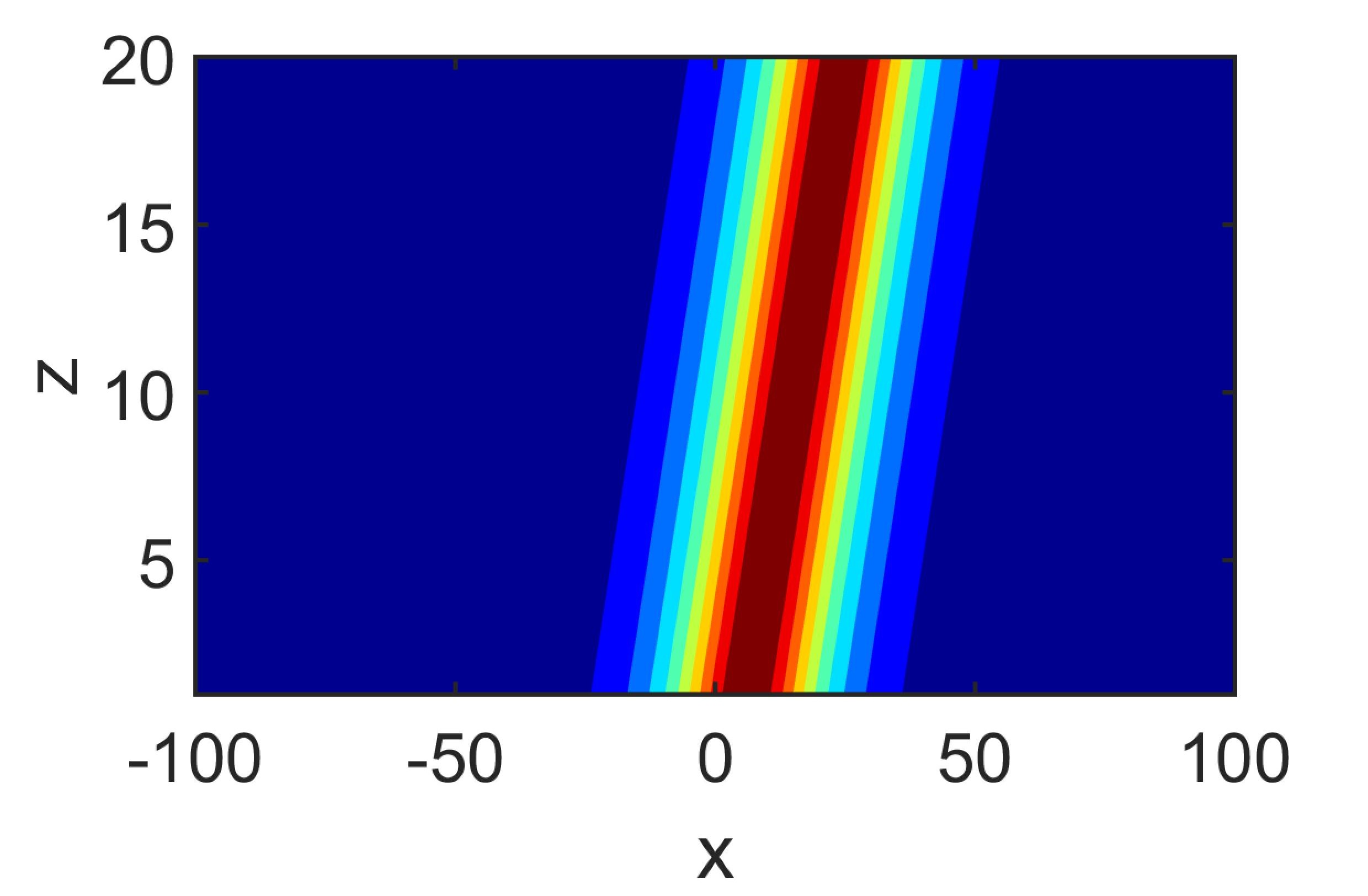}
\includegraphics[height=3.5cm]{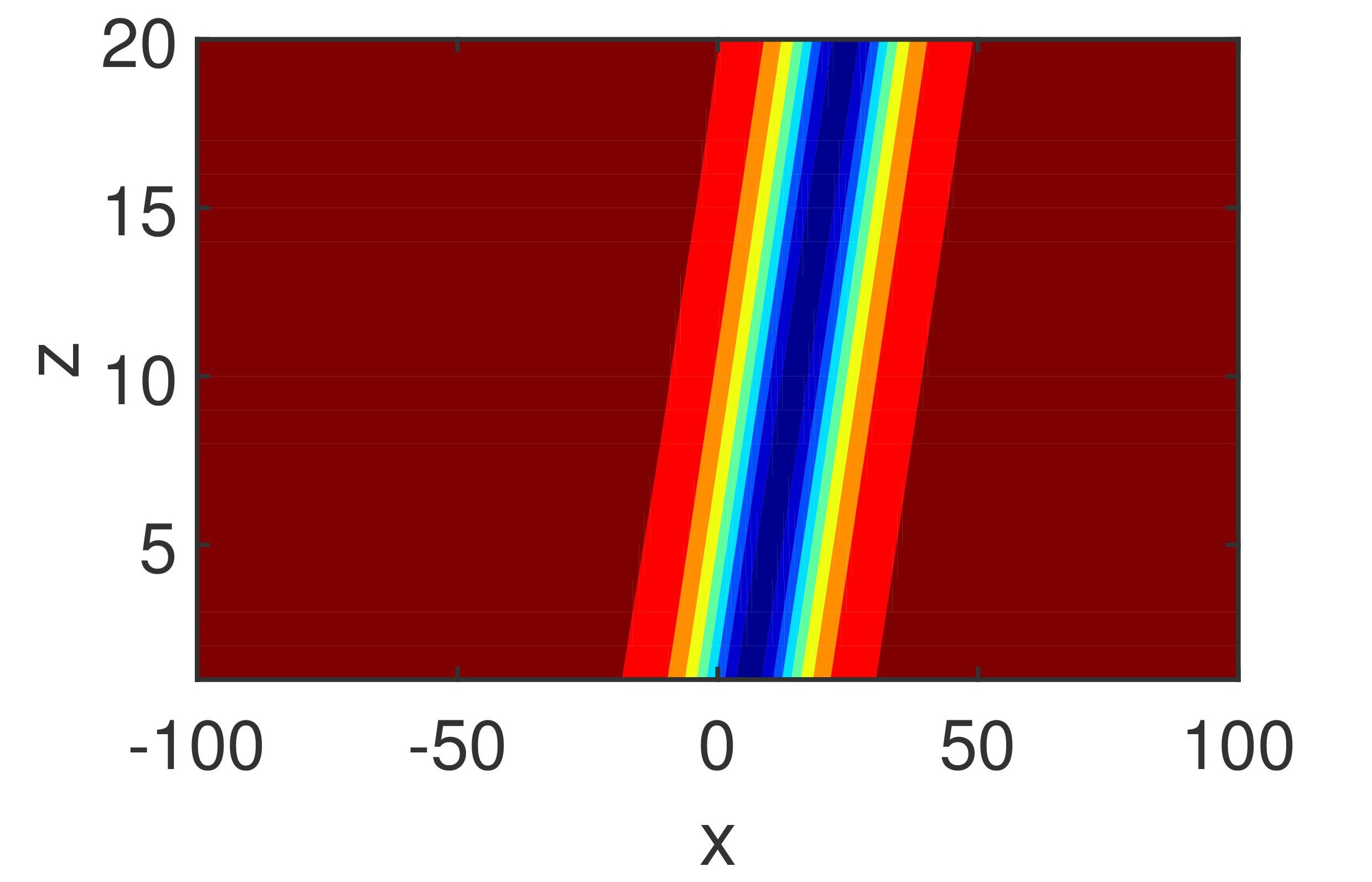}
\caption{(Color Online) Similar to Fig.~\ref{dark}, but now for a typical dark-bright
soliton pair.}
\label{darkbright}
\end{figure*}

Apart from soliton pairs of the same type, it is also possible to derive vector soliton solutions
composed by different types of solitons. This can be done upon seeking solutions of the system of
Eqs.~\eqref{system1} such that one of the components decays to zero at infinity, while the other
tends to a constant, as before. In such a case, solutions of Eqs.~\eqref{system1} are again taken
to be of the form of Eqs.~\eqref{eq.background}, but now we assume that the background functions
are given by:
\begin{subequations}
\begin{align}
u_b(x,z) &= \exp\left[ikx - i(\omega-\eps^2\Omega) z\right], \\
\omega  &= \frac{1}{2q}({d_1}{k^2}q +
4{g_1}{g_2}v_0^2), \\
{v_b}(z) &= {v_0}{\exp(- 2i{g_2}{\theta _b}z + i{\psi_1}}), \; {\theta _b} =
\frac{{{g_2}v_0^2}}{q}.
\end{align}
\end{subequations}
Then, the system~\eqref{system1} is reduced to the form:
\begin{gather}
i{u_z} + \frac{{{d_1}}}{2}{u_{xx}} - 2{g_1}{\theta _b}(w - 1)u - i{d_1}k{u_x} = 0, \\
i{v_z} + \frac{{{d_2}}}{2}{v_{xx}} - 2{g_2}{\theta _b}(w - 1)v = 0, \\
\nu {w_{xx}} - 2qw =  - \frac{2}{{{\theta _b}}}({g_1}|u{|^2} + {g_2}v_0^2|v{|^2}).
\end{gather}
Then, using the stretched variables~(\ref{str}) and the asymptotic
expansions~(\ref{as}),
and following the procedure of the previous Section, we obtain the following results.
First, at the leading order, $O(1)$, we get $\rho_{10}=0$ and $\rho_{20}=w_0=1$, while
in the linear limit, i.e., at the orders $O(\eps^2)$ and $O(\eps^3)$, we derive
equations connecting the unknown fields, namely:
%
%\[
%\rho_{10}=0,\quad \rho_{20}=w_0=1
%\]
%%
%at $O(\eps^2)$
%
\begin{subequations}
\begin{gather}
{w_2} = 2{\rho _{22}}, \quad
C\frac{{\partial {\phi _{21}}}}{{\partial X}} = \frac{{4g_2^2v_0^2}}{q}{\rho _{22}},
\\
  \frac{{{d_2}}}{2}\frac{{{\partial ^2}{\phi _{21}}}}{{\partial {X^2}}} =
  C\frac{{\partial {\rho _{22}}}}{{\partial X}}, \quad k = \frac{C}{{{d_1}}}.
%  \\ C = {d_1}k \Leftrightarrow k = \frac{C}{{{d_1}}}
\end{gather}
\end{subequations}
The above equations suggest that, now, the speed of sound is given by:
\begin{equation}
C^2=\frac{2g_2^2v_0^2d_2}{q}.
\end{equation}
Next, in the nonlinear regime, namely at $O(\eps^4)$ and $O(\eps^5)$, we obtain
the following system for the fields $\rho_{j2}$:
\begin{subequations}
\begin{gather}
  \frac{{8g_2^2v_0^2}}{{Cq}}\frac{{\partial {\rho _{22}}}}{{\partial Z}} -
  \frac{{\left( {{d_2}{q^2} - 4g_2^2v_0^2\nu }
  \right)}}{{2{q^2}}}\frac{{{\partial ^3}{\rho _{22}}}}{{\partial {X^3}}} +
  \frac{{24g_2^2v_0^2}}{q}{\rho
  _{22}}\frac{{\partial {\rho _{22}}}}{{\partial X}} \nonumber\\ +
  \frac{{2{g_1}{g_2}}}{q}\frac{\partial}{{\partial X}}\left(\rho_{12}^2\right)=0,
%  {\rho _{12}}\frac{{\partial {\rho_{12}}}}{{\partial X}} = 0,
  \\
  \frac{{{d_1}}}{2}\frac{{{\partial ^2}{\rho _{12}}}}{{\partial {X^2}}} -
  \frac{{4{g_1}{g_2}v_0^2}}{q}{\rho_{12}}{\rho
  _{22}}=\Omega \rho _{12},
\end{gather}
\label{melni}
\end{subequations}
as well as equations connecting fields that can be determined at a higher-order
approximation.
%with
%%
%\[
%{g_2}qv_0^2{w_4} = {g_1}q\rho _{12}^2 + {g_2}qv_0^2\rho _{22}^2 + 2{g_2}qv_0^2{\rho
%_{24}} + {g_2}\nu v_0^2\frac{{{\partial
%^2}{\rho _{22}}}}{{\partial {X^2}}}
%\]
%
The system of Eqs.~(\ref{melni}) is the so-called Mel’nikov system
\cite{mel1,mel2,mel3},
and is apparently composed of a KdV equation with a self-consistent source, which
satisfies
a stationary Schr{\"o}dinger equation. This system has been derived in earlier works
to
describe dark-bright solitons in nonlinear optical systems \cite{dmel1} and in
Bose-Einstein
condensates \cite{dmel2,dmel3}. The Mel'nikov system is completely integrable
by the inverse scattering transform, and possesses a soliton solution of the form
\cite{mel2}:
\begin{eqnarray}
\!\!\!\!\!\!\!\!
  {\rho _{22}}(Z,X) &=& - \frac{{{d_1}q}}{{4{g_1}{g_2}v_0^2}}{\eta
  ^2}\mathrm{sech}^2(\eta X + bZ + {X_0}),
  \\
  \!\!\!\!\!\!\!\!
  {\rho _{12}}(Z,X) &=& A\mathrm{sech} (\eta X + bZ + {X_0}),
\end{eqnarray}
where $\Omega  = (1/2)\eta^2 d_1$, while parameters $\eta$, $A$, and $b$ are connected
through the
following equation:
\begin{gather}
C{d_1}\left( {4\nu g_2^2v_0^2 - {d_2}{q^2}} \right){\eta ^4}
 + 4q{d_1}g_2^2v_0^2b\eta - 4Cg_1^2g_2^2v_0^2 A^2 = 0.
\end{gather}
Using the above expressions, we can now express the relevant approximate
[valid up to $O(\eps^2)$] solutions of the original system
for the two components $E_{1,2}$ as follows:
\begin{eqnarray}
\!\!\!\!\!\!\!\!\!\!\!\!
E_1(z,x) &\approx& \eps^2 u_b(z) \rho_{12} \exp(i\eps\phi_{12})
\label{orsol1b} \\
\!\!\!\!\!\!\!\!\!\!\!\!
E_2(z,x) &\approx& v_b(z) \left(1+\eps^2\rho_{22}\right)
\exp\left(i\eps\phi_{22}\right).
\label{orsol2b}
\end{eqnarray}
It is clear that the above solution represents a dark-bright soliton pair,
for the components $E_2$ and $E_1$, respectively.

As in the case of the dark and antidark soliton pairs, we numerically integrate
Eqs.~\eqref{system1}, using initial conditions (at $z=0$) taken from
Eqs.~(\ref{orsol1b})-(\ref{orsol2b}). The results are shown in Fig.~\ref{darkbright},
where a typical evolution of a dark-bright soliton pair is depicted.
Here we choose all parameters equal to unity, except $v_0=1/2$.
In this case too, the dark-bright soliton, not only exist, but also propagates
undistorted with constant velocity, in excellent agreement with our analytical
predictions.

\section{Conclusions}

Concluding, in this work we have completed the analysis started in Ref. \cite{tph_pla} for the
coupled focusing nonlocal NLS system. As such we studied vector nematicons in the defocusing regime
using multiscale expansion methods to derive various types of such vector solitons.

In particular, first we have found dark-dark and antidark-antidark solitons. These structures,
which have respectively the form of propagating dips or humps on top of a stable vectorial
continuous-wave background, respectively, were found to obey an effective KdV equation. The
existence of the dark or the antidark soliton pair was connected with the magnitude of the
nonlocality parameter: it was found that below (above) a certain critical value of this parameter
-- which depends on the parameters of the system, as well as the background amplitudes -- the
soliton pair is dark (antidark), much like the single nematicon system
\cite{small_amplitude}. In addition, we have found dark-bright soliton pairs, namely a dark
soliton in one component, coupled with a bright soliton in the other component. It was shown that
this soliton pair obeys another completely integrable effective model, namely the so-called
Mel'nikov system. In all cases, we have numerically integrated the original nonlocal system and
verified the existence and robustness of the vector nematicons, in excellent agreement with our
analytical approach.

It would be interesting to extend our considerations in higher-dimensional settings, and
investigate existence, stability and dynamics of such vector solitons, as well as other localized
structures, such as vortices, and combinations thereof. However, the extension of the
nematicon equations from $(1+1)$ to $(2+1)$ dimensions is a non-trivial extension. The stability
will depend on the value of $\nu$ with the solitary waves or vortices becoming unstable if $\nu$ is
small enough. This is because the nematicon equations become the $(2+1)$ dimensional NLS equation
as $\nu\rightarrow 0$.

\appendix
\section{Details on the perturbation method}

Substituting Eqs.~(\ref{str}) and (\ref{as}) into Eqs.~\eqref{madel}, we obtain
at $O(\varepsilon^2)$ the following equations:
\begin{gather}
  C\frac{{\partial {\phi _{11}}}}{{\partial X}} = \frac{{4{g_1}}}{q}({g_1}u_0^2{\rho
  _{21}} + {g_2}v_0^2{\rho _{22}}), \\
  C\frac{{\partial {\phi _{21}}}}{{\partial X}} = \frac{{4{g_2}}}{q}({g_1}u_0^2{\rho
  _{21}} + {g_2}v_0^2{\rho _{22}}), \\
  {w_2} = \frac{2}{{q{\theta _b}}}({g_1}u_0^2{\rho _{21}} + {g_2}v_0^2{\rho _{22}}),
\end{gather}
which clearly suggest that
\begin{equation}
g_2\frac{\partial \phi_{11}}{\partial X}=g_1\frac{\partial \phi_{21}}{\partial X}
\Rightarrow
\phi_{21}=\frac{g_2}{g_1}\phi_{11}.
\label{phases}
\end{equation}
Notice that any integrating constants are set to zero in order for the boundary
conditions to be
satisfied; recall in this formulation the boundary conditions are fixed at infinity.
In addition, at $O(\varepsilon^2)$ we obtain:
\begin{gather}
  \frac{{{d_1}}}{2}\frac{{{\partial ^2}{\phi _{11}}}}{{\partial {X^2}}} =
  C\frac{{\partial {\rho _{12}}}}{{\partial X}}, \\
  \frac{{{d_2}}}{2}\frac{{{\partial ^2}{\phi _{21}}}}{{\partial {X^2}}} =
  C\frac{{\partial {\rho _{22}}}}{{\partial X}},
\end{gather}
which also suggest that
\begin{equation}\label{amps}
  d_2g_2\frac{\partial\rho_{21}}{\partial X}=d_1g_1\frac{\partial\rho_{22}}{\partial
  X}
  \Rightarrow
  \rho_{22}=\frac{d_2g_2}{d_1g_1}\rho_{21},
\end{equation}
where again integrating constants have been ignored in order for the boundary
conditions to be satisfied.
%This equation suggests that the amplitudes
%of the relative solutions are proportional to each other, and hence only one equation
%will suffice to determine either $\rho_{22}$ or $\rho_{21}$.
%
Obviously, the compatibility condition of the equations yields the speed of
sound~(\ref{velocity}).
%
%\begin{equation}%\label{velocity}
%C^2 = \frac{2}{q}(d_1g_1u_0^2 + d_2g_2v_0^2).
%\end{equation}
%
The same procedure follows for the higher order equations.
When applying the above, and since $\rho_{21}$ and $\rho_{11}$
are related, one expects to find a single equation for one of the two. Hence, at
$O(\varepsilon^4)$ we derive the equations:
\begin{align}
(d_1^2g_1^2{q^2}{\theta _b}){w_4} &= (d_1^2g_1^3qu_0^2 + d_2^2g_2^3qv_0^2)\rho _{12}^2
\nonumber \\ &+ 2d_1^2g_1^3qu_0^2{\rho _{14}}
+ 2d_1^2g_1^2{g_2}qv_0^2{\rho _{24}} \nonumber \\ &+ {d_1}{g_1}\nu ({d_1}g_1^2u_0^2 +
{d_2}g_2^2v_0^2)\frac{{{\partial ^2}{\rho
_{12}}}}{{\partial {X^2}}},
\end{align}
\begin{gather}
 - 2{g_1}{\theta _b}({w_4} + {w_2}{\rho _{12}} - {\rho _{14}} + {\rho _{14}}) + C{\rho
 _{12}}\frac{{\partial {\phi
 _{11}}}}{{\partial X}} \nonumber \\- \frac{1}{2}{d_1}{\left( {\frac{{\partial {\phi
 _{11}}}}{{\partial X}}} \right)^2} +
 C\frac{{\partial {\phi _{13}}}}{{\partial X}} + \frac{1}{2}{d_1}\frac{{{\partial
 ^2}{\rho _{12}}}}{{\partial {X^2}}} -
 \frac{{\partial {\phi _{11}}}}{{\partial Z}} = 0,
\end{gather}
and
\begin{gather}
\frac{{{\partial ^2}{\phi _{13}}}}{{\partial {X^2}}} =  - \frac{2}{{{d_1}}}\left(
{3C{\rho _{12}}\frac{{\partial {\rho
_{12}}}}{{\partial X}} - C\frac{{\partial {\rho _{14}}}}{{\partial X}} +
\frac{{\partial {\rho _{12}}}}{{\partial Z}}}
\right),
\end{gather}
while at $O(\varepsilon^5)$ we obtain:
\begin{gather}
(8d_1^2g_1^2g_2^2qv_0^2){\rho _{24}} = - 8d{1^2}g{1^3}g2qu{0^2}{\rho _{14}} \nonumber
\\-
4{d_1}{g_1}{g_2}q({d_1}g_1^2u_0^2 +
2{d_2}{g_1}{g_2}u_0^2 + 3{d_1}g_2^3v_0^2)\rho _{12}^2 \nonumber \\-
{d_1}{d_2}{g_1}{g_2}(4{d_1}g_1^2\nu u_0^2 + 4g_2^2\nu v_0^2 -
{d_2}{q^2})\frac{{{\partial ^2}{\rho _{12}}}}{{\partial {X^2}}} \nonumber \\+
2Cd_1^2g_1^2{q^2}\frac{{\partial {\phi
_{23}}}}{{\partial X}} - 2d_1^2{g_1}{g_2}{q^2}\frac{{\partial {\phi _{11}}}}{{\partial
Z}},
\end{gather}
\begin{gather}
\frac{{2d_1^2g_1^2q\left( {{C^2}q - 2{d_2}g_2^2v_0^2}
\right)}}{{{g_2}}}\frac{{{\partial ^2}{\phi _{23}}}}{{\partial
{X^2}}}=  8Cd_1^2g_1^3qu_0^2\frac{{\partial {\rho _{14}}}}{{\partial X}}
\nonumber \\+8C{g_1}q(d_1^2g_1^2u_0^2 +
2{d_1}{d_2}{g_1}{g_2}u_0^2 + 6d_2^2g_2^3v_0^2){\rho _{12}}\frac{{\partial {\rho
_{12}}}}{{\partial X}}  \nonumber \\+ C{d_1}{g_1}( -
d_2^2{q^2} + 4{d_1}g_1^2u_0^2\nu  + 4{d_2}g_2^2v_0^2\nu )\frac{{{\partial ^3}{\rho
_{12}}}}{{\partial {X^3}}} \nonumber \\+
4{d_1}{g_1}q({C^2}q + 2{d_2}g_2^2v_0^2)\frac{{\partial {\rho _{12}}}}{{\partial Z}}.
\end{gather}
To this end, after a tedious but straightforward calculation, we eliminate all phase
terms from these systems, and derive the KdV equation (\ref{kdv}).

%\bibliography{biblio_nematicons}

\begin{thebibliography}{46}
\expandafter\ifx\csname natexlab\endcsname\relax\def\natexlab#1{#1}\fi
\expandafter\ifx\csname bibnamefont\endcsname\relax
  \def\bibnamefont#1{#1}\fi
\expandafter\ifx\csname bibfnamefont\endcsname\relax
  \def\bibfnamefont#1{#1}\fi
\expandafter\ifx\csname citenamefont\endcsname\relax
  \def\citenamefont#1{#1}\fi
\expandafter\ifx\csname url\endcsname\relax
  \def\url#1{\texttt{#1}}\fi
\expandafter\ifx\csname urlprefix\endcsname\relax\def\urlprefix{URL }\fi
\providecommand{\bibinfo}[2]{#2}
\providecommand{\eprint}[2][]{\url{#2}}

\bibitem[{\citenamefont{Dauxois and Peyrard}(2006)}]{daux}
\bibinfo{author}{\bibfnamefont{T.}~\bibnamefont{Dauxois}} \bibnamefont{and}
  \bibinfo{author}{\bibfnamefont{M.}~\bibnamefont{Peyrard}},
  \emph{\bibinfo{title}{Physics of Solitons}} (\bibinfo{publisher}{Cambridge
  University Press}, \bibinfo{year}{2006}).

\bibitem[{\citenamefont{Kivshar and Agrawal}(2003)}]{kivshar_book}
\bibinfo{author}{\bibfnamefont{Y.~S.} \bibnamefont{Kivshar}} \bibnamefont{and}
  \bibinfo{author}{\bibfnamefont{G.~P.} \bibnamefont{Agrawal}},
  \emph{\bibinfo{title}{Optical Solitons: From Fibers to Photonic Crystals}}
  (\bibinfo{publisher}{Academic Press}, \bibinfo{year}{2003}).

\bibitem[{\citenamefont{Kivshar and Luther-Davies}(1998)}]{ofy}
\bibinfo{author}{\bibfnamefont{Y.~S.} \bibnamefont{Kivshar}} \bibnamefont{and}
  \bibinfo{author}{\bibfnamefont{B.}~\bibnamefont{Luther-Davies}},
  \bibinfo{journal}{Phys. Rep.} \textbf{\bibinfo{volume}{298}},
  \bibinfo{pages}{81} (\bibinfo{year}{1998}).

\bibitem[{\citenamefont{Kevrekidis et~al.}(2015)\citenamefont{Kevrekidis,
  Frantzeskakis, and Carretero-Gonz\'alez}}]{siam}
\bibinfo{author}{\bibfnamefont{P.~G.} \bibnamefont{Kevrekidis}},
  \bibinfo{author}{\bibfnamefont{D.~J.} \bibnamefont{Frantzeskakis}},
  \bibnamefont{and}
  \bibinfo{author}{\bibfnamefont{R.}~\bibnamefont{Carretero-Gonz\'alez}},
  \emph{\bibinfo{title}{The defocusing nonlinear Schr\"odinger equation: from
  dark solitons to vortices and vortex rings}} (\bibinfo{publisher}{SIAM},
  \bibinfo{year}{2015}).

\bibitem[{\citenamefont{Kevrekidis and Frantzeskakis}(2016)}]{revip}
\bibinfo{author}{\bibfnamefont{P.~G.} \bibnamefont{Kevrekidis}}
  \bibnamefont{and} \bibinfo{author}{\bibfnamefont{D.~J.}
  \bibnamefont{Frantzeskakis}}, \bibinfo{journal}{Rev. in Phys.}
  \textbf{\bibinfo{volume}{1}}, \bibinfo{pages}{140} (\bibinfo{year}{2016}).

\bibitem[{\citenamefont{Litvak et~al.}(1975)\citenamefont{Litvak, Mironov,
  Fraiman, and Yunakovskii}}]{litvak}
\bibinfo{author}{\bibfnamefont{A.~G.} \bibnamefont{Litvak}},
  \bibinfo{author}{\bibfnamefont{V.~A.} \bibnamefont{Mironov}},
  \bibinfo{author}{\bibfnamefont{G.~M.} \bibnamefont{Fraiman}},
  \bibnamefont{and} \bibinfo{author}{\bibfnamefont{A.~D.}
  \bibnamefont{Yunakovskii}}, \bibinfo{journal}{Sov. J. Plasma Phys.}
  \textbf{\bibinfo{volume}{1}}, \bibinfo{pages}{60} (\bibinfo{year}{1975}).

\bibitem[{\citenamefont{Yakimenko et~al.}(2005)\citenamefont{Yakimenko,
  Zaliznyak, and Kivshar}}]{plasma}
\bibinfo{author}{\bibfnamefont{A.~I.} \bibnamefont{Yakimenko}},
  \bibinfo{author}{\bibfnamefont{Y.~A.} \bibnamefont{Zaliznyak}},
  \bibnamefont{and} \bibinfo{author}{\bibfnamefont{Y.~S.}
  \bibnamefont{Kivshar}}, \bibinfo{journal}{Phys. Rev. E}
  \textbf{\bibinfo{volume}{71}}, \bibinfo{pages}{065603(R)}
  (\bibinfo{year}{2005}).

\bibitem[{\citenamefont{Suter and Blasberg}(1993)}]{vap}
\bibinfo{author}{\bibfnamefont{D.}~\bibnamefont{Suter}} \bibnamefont{and}
  \bibinfo{author}{\bibfnamefont{T.}~\bibnamefont{Blasberg}},
  \bibinfo{journal}{Phys. Rev. A} \textbf{\bibinfo{volume}{48}},
  \bibinfo{pages}{4583} (\bibinfo{year}{1993}).

\bibitem[{\citenamefont{Rotschild et~al.}(2005)\citenamefont{Rotschild, Cohen,
  Manela, Segev, and Carmon}}]{rot}
\bibinfo{author}{\bibfnamefont{C.}~\bibnamefont{Rotschild}},
  \bibinfo{author}{\bibfnamefont{O.}~\bibnamefont{Cohen}},
  \bibinfo{author}{\bibfnamefont{O.}~\bibnamefont{Manela}},
  \bibinfo{author}{\bibfnamefont{M.}~\bibnamefont{Segev}}, \bibnamefont{and}
  \bibinfo{author}{\bibfnamefont{T.}~\bibnamefont{Carmon}},
  \bibinfo{journal}{Phys. Rev. Lett.} \textbf{\bibinfo{volume}{95}},
  \bibinfo{pages}{213904} (\bibinfo{year}{2005}).

\bibitem[{\citenamefont{Pedri and Santos}(2005)}]{santos}
\bibinfo{author}{\bibfnamefont{P.}~\bibnamefont{Pedri}} \bibnamefont{and}
  \bibinfo{author}{\bibfnamefont{L.}~\bibnamefont{Santos}},
  \bibinfo{journal}{Phys. Rev. Lett.} \textbf{\bibinfo{volume}{95}},
  \bibinfo{pages}{200404} (\bibinfo{year}{2005}).

\bibitem[{\citenamefont{Conti et~al.}(2003)\citenamefont{Conti, Peccianti, and
  Assanto}}]{ass0}
\bibinfo{author}{\bibfnamefont{C.}~\bibnamefont{Conti}},
  \bibinfo{author}{\bibfnamefont{M.}~\bibnamefont{Peccianti}},
  \bibnamefont{and} \bibinfo{author}{\bibfnamefont{G.}~\bibnamefont{Assanto}},
  \bibinfo{journal}{Phys. Rev. Lett.} \textbf{\bibinfo{volume}{91}},
  \bibinfo{pages}{073901} (\bibinfo{year}{2003}).

\bibitem[{\citenamefont{Assanto and Peccianti}(2003)}]{assanto}
\bibinfo{author}{\bibfnamefont{G.}~\bibnamefont{Assanto}} \bibnamefont{and}
  \bibinfo{author}{\bibfnamefont{M.}~\bibnamefont{Peccianti}},
  \bibinfo{journal}{IEEE J. Quant. Electr.} \textbf{\bibinfo{volume}{39}},
  \bibinfo{pages}{13} (\bibinfo{year}{2003}).

\bibitem[{\citenamefont{Assanto
  et~al.}(2009{\natexlab{a}})\citenamefont{Assanto, Minzoni, and Smyth}}]{ass2}
\bibinfo{author}{\bibfnamefont{G.}~\bibnamefont{Assanto}},
  \bibinfo{author}{\bibfnamefont{A.~A.} \bibnamefont{Minzoni}},
  \bibnamefont{and} \bibinfo{author}{\bibfnamefont{N.~F.} \bibnamefont{Smyth}},
  \bibinfo{journal}{J. Nonlinear Opt. Phys. Mater.}
  \textbf{\bibinfo{volume}{18}}, \bibinfo{pages}{657}
  (\bibinfo{year}{2009}{\natexlab{a}}).

\bibitem[{\citenamefont{Assanto}(2012)}]{assanto_book}
\bibinfo{author}{\bibfnamefont{G.}~\bibnamefont{Assanto}},
  \emph{\bibinfo{title}{Nematicons: Spatial Optical Solitons in Nematic Liquid
  Crystals}} (\bibinfo{publisher}{Wiley-Blackwell}, \bibinfo{year}{2012}).

\bibitem[{\citenamefont{Peccianti and Assanto}(2012)}]{peccianti}
\bibinfo{author}{\bibfnamefont{M.}~\bibnamefont{Peccianti}} \bibnamefont{and}
  \bibinfo{author}{\bibfnamefont{G.}~\bibnamefont{Assanto}},
  \bibinfo{journal}{Phys. Rep.} \textbf{\bibinfo{volume}{516}},
  \bibinfo{pages}{147} (\bibinfo{year}{2012}).

\bibitem[{\citenamefont{Minzoni et~al.}(2007)\citenamefont{Minzoni, Smyth, and
  Worthy}}]{worthy2}
\bibinfo{author}{\bibfnamefont{A.~A.} \bibnamefont{Minzoni}},
  \bibinfo{author}{\bibfnamefont{N.~F.} \bibnamefont{Smyth}}, \bibnamefont{and}
  \bibinfo{author}{\bibfnamefont{A.~L.} \bibnamefont{Worthy}},
  \bibinfo{journal}{J. Opt. Soc. Am. B} \textbf{\bibinfo{volume}{24}},
  \bibinfo{pages}{1549} (\bibinfo{year}{2007}).

\bibitem[{\citenamefont{Skuse and Smyth}(2008)}]{skuse3}
\bibinfo{author}{\bibfnamefont{B.~D.} \bibnamefont{Skuse}} \bibnamefont{and}
  \bibinfo{author}{\bibfnamefont{N.~F.} \bibnamefont{Smyth}},
  \bibinfo{journal}{Phys. Rev. A} \textbf{\bibinfo{volume}{77}},
  \bibinfo{pages}{013817} (\bibinfo{year}{2008}).

\bibitem[{\citenamefont{Alberucci and Assanto}(2013)}]{alberucci}
\bibinfo{author}{\bibfnamefont{A.}~\bibnamefont{Alberucci}} \bibnamefont{and}
  \bibinfo{author}{\bibfnamefont{G.}~\bibnamefont{Assanto}},
  \bibinfo{journal}{Mol. Cryst. Liq. Cryst.} \textbf{\bibinfo{volume}{572}},
  \bibinfo{pages}{2} (\bibinfo{year}{2013}).

\bibitem[{\citenamefont{Assanto
  et~al.}(2009{\natexlab{b}})\citenamefont{Assanto, Minzoni, and
  Smyth}}]{assanto3}
\bibinfo{author}{\bibfnamefont{G.}~\bibnamefont{Assanto}},
  \bibinfo{author}{\bibfnamefont{A.~A.} \bibnamefont{Minzoni}},
  \bibnamefont{and} \bibinfo{author}{\bibfnamefont{N.~F.} \bibnamefont{Smyth}},
  \bibinfo{journal}{J. Nonlinear Opt. Phys. Mater.}
  \textbf{\bibinfo{volume}{18}}, \bibinfo{pages}{657}
  (\bibinfo{year}{2009}{\natexlab{b}}).

\bibitem[{\citenamefont{Kartashov and Torner}(2007)}]{karta}
\bibinfo{author}{\bibfnamefont{Y.~V.} \bibnamefont{Kartashov}}
  \bibnamefont{and} \bibinfo{author}{\bibfnamefont{L.}~\bibnamefont{Torner}},
  \bibinfo{journal}{Opt. Lett.} \textbf{\bibinfo{volume}{32}},
  \bibinfo{pages}{946–948} (\bibinfo{year}{2007}).

\bibitem[{\citenamefont{Piccardi et~al.}(2011)\citenamefont{Piccardi,
  Alberucci, Tabiryan, and Assanto}}]{piccardi}
\bibinfo{author}{\bibfnamefont{A.}~\bibnamefont{Piccardi}},
  \bibinfo{author}{\bibfnamefont{A.}~\bibnamefont{Alberucci}},
  \bibinfo{author}{\bibfnamefont{N.}~\bibnamefont{Tabiryan}}, \bibnamefont{and}
  \bibinfo{author}{\bibfnamefont{G.}~\bibnamefont{Assanto}},
  \bibinfo{journal}{Opt. Lett.} \textbf{\bibinfo{volume}{36}},
  \bibinfo{pages}{1356} (\bibinfo{year}{2011}).

\bibitem[{\citenamefont{Kong et~al.}(2010)\citenamefont{Kong, Wang, Bang, and
  Kr\'olikowski}}]{kong}
\bibinfo{author}{\bibfnamefont{Q.}~\bibnamefont{Kong}},
  \bibinfo{author}{\bibfnamefont{Q.}~\bibnamefont{Wang}},
  \bibinfo{author}{\bibfnamefont{O.}~\bibnamefont{Bang}}, \bibnamefont{and}
  \bibinfo{author}{\bibfnamefont{W.}~\bibnamefont{Kr\'olikowski}},
  \bibinfo{journal}{Opt. Lett.} \textbf{\bibinfo{volume}{35}},
  \bibinfo{pages}{2152} (\bibinfo{year}{2010}).

\bibitem[{\citenamefont{Assanto et~al.}(2011)\citenamefont{Assanto, Marchant,
  Minzoni, and Smyth}}]{assanto_grey}
\bibinfo{author}{\bibfnamefont{G.}~\bibnamefont{Assanto}},
  \bibinfo{author}{\bibfnamefont{T.~R.} \bibnamefont{Marchant}},
  \bibinfo{author}{\bibfnamefont{A.~A.} \bibnamefont{Minzoni}},
  \bibnamefont{and} \bibinfo{author}{\bibfnamefont{N.~F.} \bibnamefont{Smyth}},
  \bibinfo{journal}{Phys. Rev. E} \textbf{\bibinfo{volume}{84}},
  \bibinfo{pages}{066602} (\bibinfo{year}{2011}).

\bibitem[{\citenamefont{Pu et~al.}(2012)\citenamefont{Pu, Hou, Zhan, and
  Yuan}}]{pu}
\bibinfo{author}{\bibfnamefont{S.}~\bibnamefont{Pu}},
  \bibinfo{author}{\bibfnamefont{C.}~\bibnamefont{Hou}},
  \bibinfo{author}{\bibfnamefont{K.}~\bibnamefont{Zhan}}, \bibnamefont{and}
  \bibinfo{author}{\bibfnamefont{C.}~\bibnamefont{Yuan}},
  \bibinfo{journal}{Physica Scr.} \textbf{\bibinfo{volume}{85}},
  \bibinfo{pages}{015402} (\bibinfo{year}{2012}).

\bibitem[{\citenamefont{Horikis}(2015)}]{small_amplitude}
\bibinfo{author}{\bibfnamefont{T.~P.} \bibnamefont{Horikis}},
  \bibinfo{journal}{J. Phys. A: Math. Theor.} \textbf{\bibinfo{volume}{48}},
  \bibinfo{pages}{02FT01} (\bibinfo{year}{2015}).

\bibitem[{\citenamefont{Horikis and Frantzeskakis}(2016{\natexlab{a}})}]{djfth}
\bibinfo{author}{\bibfnamefont{T.~P.} \bibnamefont{Horikis}} \bibnamefont{and}
  \bibinfo{author}{\bibfnamefont{D.~J.} \bibnamefont{Frantzeskakis}},
  \bibinfo{journal}{Opt. Lett.} \textbf{\bibinfo{volume}{41}},
  \bibinfo{pages}{583} (\bibinfo{year}{2016}{\natexlab{a}}).

\bibitem[{\citenamefont{Horikis and
  Frantzeskakis}(2016{\natexlab{b}})}]{djfth2}
\bibinfo{author}{\bibfnamefont{T.~P.} \bibnamefont{Horikis}} \bibnamefont{and}
  \bibinfo{author}{\bibfnamefont{D.~J.} \bibnamefont{Frantzeskakis}},
  \bibinfo{journal}{J. Phys. A: Math. Theor.} \textbf{\bibinfo{volume}{49}},
  \bibinfo{pages}{205202} (\bibinfo{year}{2016}{\natexlab{b}}).

\bibitem[{\citenamefont{Alberucci et~al.}(2006)\citenamefont{Alberucci,
  Peccianti, Assanto, Dyadyusha, and Kaczmarek}}]{vector}
\bibinfo{author}{\bibfnamefont{A.}~\bibnamefont{Alberucci}},
  \bibinfo{author}{\bibfnamefont{M.}~\bibnamefont{Peccianti}},
  \bibinfo{author}{\bibfnamefont{G.}~\bibnamefont{Assanto}},
  \bibinfo{author}{\bibfnamefont{A.}~\bibnamefont{Dyadyusha}},
  \bibnamefont{and}
  \bibinfo{author}{\bibfnamefont{M.}~\bibnamefont{Kaczmarek}},
  \bibinfo{journal}{Phys. Rev. Lett.} \textbf{\bibinfo{volume}{97}},
  \bibinfo{pages}{153903} (\bibinfo{year}{2006}).

\bibitem[{\citenamefont{Assanto et~al.}(2008)\citenamefont{Assanto, Smyth, and
  Worthy}}]{worthy}
\bibinfo{author}{\bibfnamefont{G.}~\bibnamefont{Assanto}},
  \bibinfo{author}{\bibfnamefont{N.~F.} \bibnamefont{Smyth}}, \bibnamefont{and}
  \bibinfo{author}{\bibfnamefont{A.~L.} \bibnamefont{Worthy}},
  \bibinfo{journal}{Phys. Rev. A} \textbf{\bibinfo{volume}{78}},
  \bibinfo{pages}{013832} (\bibinfo{year}{2008}).

\bibitem[{\citenamefont{Xu et~al.}(2009)\citenamefont{Xu, Smyth, Minzoni, and
  Kivshar}}]{xu}
\bibinfo{author}{\bibfnamefont{Z.}~\bibnamefont{Xu}},
  \bibinfo{author}{\bibfnamefont{N.~F.} \bibnamefont{Smyth}},
  \bibinfo{author}{\bibfnamefont{A.~A.} \bibnamefont{Minzoni}},
  \bibnamefont{and} \bibinfo{author}{\bibfnamefont{Y.~S.}
  \bibnamefont{Kivshar}}, \bibinfo{journal}{Opt. Lett.}
  \textbf{\bibinfo{volume}{34}}, \bibinfo{pages}{1414} (\bibinfo{year}{2009}).

\bibitem[{\citenamefont{Horikis}(2016)}]{tph_pla}
\bibinfo{author}{\bibfnamefont{T.~P.} \bibnamefont{Horikis}},
  \bibinfo{journal}{Phys. Lett. A} \textbf{\bibinfo{volume}{380}},
  \bibinfo{pages}{3473} (\bibinfo{year}{2016}).

\bibitem[{\citenamefont{Chen et~al.}(2013)\citenamefont{Chen, Kong, Shen, Wang,
  and Shi}}]{kong1}
\bibinfo{author}{\bibfnamefont{W.}~\bibnamefont{Chen}},
  \bibinfo{author}{\bibfnamefont{Q.}~\bibnamefont{Kong}},
  \bibinfo{author}{\bibfnamefont{M.}~\bibnamefont{Shen}},
  \bibinfo{author}{\bibfnamefont{Q.}~\bibnamefont{Wang}}, \bibnamefont{and}
  \bibinfo{author}{\bibfnamefont{J.}~\bibnamefont{Shi}},
  \bibinfo{journal}{Phys. Rev. E} \textbf{\bibinfo{volume}{87}},
  \bibinfo{pages}{013809} (\bibinfo{year}{2013}).

\bibitem[{\citenamefont{Lin and Lee}(2007)}]{lin1}
\bibinfo{author}{\bibfnamefont{Y.}~\bibnamefont{Lin}} \bibnamefont{and}
  \bibinfo{author}{\bibfnamefont{R.-K.} \bibnamefont{Lee}},
  \bibinfo{journal}{Opt. Express} \textbf{\bibinfo{volume}{15}},
  \bibinfo{pages}{8781} (\bibinfo{year}{2007}).

\bibitem[{\citenamefont{Khoo et~al.}(1993)\citenamefont{Khoo, Li, and
  Liang}}]{khoo}
\bibinfo{author}{\bibfnamefont{I.}~\bibnamefont{Khoo}},
  \bibinfo{author}{\bibfnamefont{H.}~\bibnamefont{Li}}, \bibnamefont{and}
  \bibinfo{author}{\bibfnamefont{Y.}~\bibnamefont{Liang}},
  \bibinfo{journal}{IEEE J. Quantum Electron.} \textbf{\bibinfo{volume}{29}},
  \bibinfo{pages}{1444–1447} (\bibinfo{year}{1993}).

\bibitem[{\citenamefont{{Mel'nikov}}(1983)}]{mel1}
\bibinfo{author}{\bibfnamefont{V.~K.} \bibnamefont{{Mel'nikov}}},
  \bibinfo{journal}{Lett. Math. Phys.} \textbf{\bibinfo{volume}{7}},
  \bibinfo{pages}{129} (\bibinfo{year}{1983}).

\bibitem[{\citenamefont{{Mel'nikov}}(1988{\natexlab{a}})}]{mel2}
\bibinfo{author}{\bibfnamefont{V.~K.} \bibnamefont{{Mel'nikov}}},
  \bibinfo{journal}{Phys. Lett. A} \textbf{\bibinfo{volume}{128}},
  \bibinfo{pages}{488} (\bibinfo{year}{1988}{\natexlab{a}}).

\bibitem[{\citenamefont{{Mel'nikov}}(1988{\natexlab{b}})}]{mel3}
\bibinfo{author}{\bibfnamefont{V.~K.} \bibnamefont{{Mel'nikov}}},
  \bibinfo{journal}{Phys. Lett. A} \textbf{\bibinfo{volume}{133}},
  \bibinfo{pages}{493} (\bibinfo{year}{1988}{\natexlab{b}}).

\bibitem[{\citenamefont{Skuse and Smyth}(2009)}]{skuse2}
\bibinfo{author}{\bibfnamefont{B.~D.} \bibnamefont{Skuse}} \bibnamefont{and}
  \bibinfo{author}{\bibfnamefont{N.~F.} \bibnamefont{Smyth}},
  \bibinfo{journal}{Phys. Rev. A} \textbf{\bibinfo{volume}{79}},
  \bibinfo{pages}{063806} (\bibinfo{year}{2009}).

\bibitem[{\citenamefont{MacNeil et~al.}(2014)\citenamefont{MacNeil, Smyth, and
  Assanto}}]{mcneil}
\bibinfo{author}{\bibfnamefont{J.~M.~L.} \bibnamefont{MacNeil}},
  \bibinfo{author}{\bibfnamefont{N.~F.} \bibnamefont{Smyth}}, \bibnamefont{and}
  \bibinfo{author}{\bibfnamefont{G.}~\bibnamefont{Assanto}},
  \bibinfo{journal}{Physica D} \textbf{\bibinfo{volume}{284}},
  \bibinfo{pages}{1} (\bibinfo{year}{2014}).

\bibitem[{\citenamefont{Jeffrey and Kawahara}(1982)}]{Jeffrey}
\bibinfo{author}{\bibfnamefont{A.}~\bibnamefont{Jeffrey}} \bibnamefont{and}
  \bibinfo{author}{\bibfnamefont{T.}~\bibnamefont{Kawahara}},
  \emph{\bibinfo{title}{Asymptotic methods in nonlinear wave theory}}
  (\bibinfo{publisher}{Pitman}, \bibinfo{year}{1982}).

\bibitem[{\citenamefont{Ablowitz and Segur}(1981)}]{ist}
\bibinfo{author}{\bibfnamefont{M.~J.} \bibnamefont{Ablowitz}} \bibnamefont{and}
  \bibinfo{author}{\bibfnamefont{H.}~\bibnamefont{Segur}},
  \emph{\bibinfo{title}{Solitons and the Inverse Scattering Transform}}
  (\bibinfo{publisher}{SIAM Studies in Applied Mathematics},
  \bibinfo{year}{1981}).

\bibitem[{\citenamefont{Ablowitz}(2011)}]{ablowitz}
\bibinfo{author}{\bibfnamefont{M.~J.} \bibnamefont{Ablowitz}},
  \emph{\bibinfo{title}{Nonlinear dispersive waves: {A}symptotic analysis and
  solitons}} (\bibinfo{publisher}{Cambridge University Press},
  \bibinfo{year}{2011}).

\bibitem[{\citenamefont{El and Smyth}(2016)}]{el}
\bibinfo{author}{\bibfnamefont{G.}~\bibnamefont{El}} \bibnamefont{and}
  \bibinfo{author}{\bibfnamefont{N.~F.} \bibnamefont{Smyth}},
  \bibinfo{journal}{Proc. Roy. Soc. Lond. A} \textbf{\bibinfo{volume}{472}},
  \bibinfo{pages}{20150633} (\bibinfo{year}{2016}).

\bibitem[{\citenamefont{Frantzeskakis}(2001)}]{dmel1}
\bibinfo{author}{\bibfnamefont{D.~J.} \bibnamefont{Frantzeskakis}},
  \bibinfo{journal}{Phys. Lett. A} \textbf{\bibinfo{volume}{285}},
  \bibinfo{pages}{363} (\bibinfo{year}{2001}).

\bibitem[{\citenamefont{Aguero et~al.}(2006)\citenamefont{Aguero,
  Frantzeskakis, and Kevrekidis}}]{dmel2}
\bibinfo{author}{\bibfnamefont{M.}~\bibnamefont{Aguero}},
  \bibinfo{author}{\bibfnamefont{D.~J.} \bibnamefont{Frantzeskakis}},
  \bibnamefont{and} \bibinfo{author}{\bibfnamefont{P.~G.}
  \bibnamefont{Kevrekidis}}, \bibinfo{journal}{J. Phys. A: Math. Gen.}
  \textbf{\bibinfo{volume}{39}}, \bibinfo{pages}{7705} (\bibinfo{year}{2006}).

\bibitem[{\citenamefont{Tsitoura et~al.}(2013)\citenamefont{Tsitoura,
  Achilleos, Malomed, Yan, Kevrekidis, and Frantzeskakis}}]{dmel3}
\bibinfo{author}{\bibfnamefont{F.}~\bibnamefont{Tsitoura}},
  \bibinfo{author}{\bibfnamefont{V.}~\bibnamefont{Achilleos}},
  \bibinfo{author}{\bibfnamefont{B.~A.} \bibnamefont{Malomed}},
  \bibinfo{author}{\bibfnamefont{D.}~\bibnamefont{Yan}},
  \bibinfo{author}{\bibfnamefont{P.~G.} \bibnamefont{Kevrekidis}},
  \bibnamefont{and} \bibinfo{author}{\bibfnamefont{D.~J.}
  \bibnamefont{Frantzeskakis}}, \bibinfo{journal}{Phys. Rev. A}
  \textbf{\bibinfo{volume}{87}}, \bibinfo{pages}{063624}
  (\bibinfo{year}{2013}).

\end{thebibliography}

\end{document}